\documentclass[11pt,a4paper]{article}
\pdfoutput=1
\usepackage{jheppub}
\usepackage{amsmath}
\usepackage{dsfont}
\usepackage{MnSymbol}
\usepackage{ulem}
\usepackage{array}
\usepackage{multirow}
\usepackage{hhline}

\renewcommand{\arraystretch}{1.4}
\makeatletter\renewcommand{\@biblabel}[1]{#1.}\makeatother

\def\be{\begin{equation}}
\def\ee{\end{equation}}

\def\p{\partial}
\def\nn{\nonumber}
\def\rd{{\rm d}}
\def\bs{\boldsymbol}
\def\i{{\rm i}}
\def\e{{\rm e}}
\def\q{{\mathfrak{q}}}
\def\p{{\mathfrak{p}}}
\def\t{{\mathfrak{t}}}
\def\a{{\sf a}}
\def\s{{\sf s}}
\def\b{{\sf b}}
\def\L{{\sf L}}
\def\V{{\sf V}}
\def\J{{\sf J}}
\def\Y{{\sf Y}}
\def\y{{\sf y}}
\def\S{{\sf S}}

\def\O{{\sf O}}

\def\T{{\sf T}}

\def\H{{\sf H}}
\def\t{{\mathfrak t}}
\def\Z{{\sf Z}}
\def\ul#1{\underline{#1}}

\def\ket#1{{ |#1\rangle}}
\def\bra#1{{ \langle#1|}}
\def\braket #1#2{{ \langle #1|#2 \rangle}}

\preprint{
{\small{\textsf{}}}}

\title{Elliptic modular double and 4d partition functions}

\author[a]{Rebecca Lodin, Fabrizio Nieri, Maxim Zabzine}

\affiliation[a]{Department of Physics and Astronomy, Uppsala University,\\
Box 516, SE-75120 Uppsala, Sweden.}

\emailAdd{rebecca.lodin@physics.uu.se}
\emailAdd{fb.nieri@gmail.com}
\emailAdd{maxim.zabzine@physics.uu.se}

\abstract{We consider 4d supersymmetric (special) unitary $\Gamma$ quiver gauge theories on compact manifolds which are $\mathbb{T}^2$ fibrations over $\mathbb{S}^2$. We show that their partition functions are correlators of vertex operators and screening charges of the modular double version of elliptic $W_{q,t;q'}(\Gamma)$ algebras. We also consider a generating function of BPS surface defects supported on $\mathbb{T}^2$ and show that it can be identified with a particular coherent state in the Fock module over the elliptic Heisenberg algebra.
 
}


\begin{document}
\maketitle

\flushbottom

\section{Introduction}
One of the major breakthroughs of the last decade was the discovery that protected sectors of certain supersymmetric gauge theories in various dimensions can be described by infinite dimensional algebras. This idea goes generically under the name of BPS/CFT correspondence (see \cite{Nekrasov:2015wsu,Nekrasov:2016qym,Nekrasov:2016ydq} for a recent review and developments). Remarkable realizations of this program are the gauge/Bethe correspondence \cite{Nekrasov:2009uh,Nekrasov:2009ui,Nekrasov:2009rc}, the chiral algebras in 4d or 6d SCFTs  \cite{Beem:2013sza,Beem:2014kka}, and the AGT correspondence \cite{Alday:2009aq,Wyllard:2009hg} which identifies the $\mathbb{S}^4$ partition function \cite{Pestun:2007rz,Hama:2012bg} of a wide class of 4d $\mathcal{N}=2$ theories (class $\mathcal{S}$) \cite{Gaiotto:2009we} with 2d CFT correlators with $W$ symmetry (Liouville/Toda theories). As a consequence, $\mathbb{R}^4$ Nekrasov partition functions \cite{Nekrasov:2002qd,Nekrasov:2003rj} can be identified with $W$ conformal blocks (see \cite{Teschner:2016yzf} for a recent review). Exploring possible extensions of the AGT duality is the main motivation for this work. 

It has been known for a long time that the AGT correspondence can be {\it $q$-deformed} \cite{Awata:2009ur,Awata:2010yy,Awata:2011dc,Mironov:2011dk} (see also \cite{Awata:2016riz,Mironov:2016yue,Awata:2016bdm,Kimura:2015rgi} for recent developments). From the gauge theory viewpoint, the deformation amounts to the 5d lifting, either on $\mathbb{S}^4\times \mathbb{S}^1$, $\mathbb{S}^5$ or $\mathbb{R}^4\times \mathbb{S}^1$ \cite{Nieri:2013yra,Nieri:2013vba,Carlsson:2013jka}, while from the 2d CFT perspective it corresponds to the {\it trigonometric deformation} of the $W$ algebra known as $W_{q,t}$ \cite{Shiraishi:1995rp,Awata:1995zk,1997q.alg.....8006F}. The $q$-deformed AGT correspondence can also be studied in a slightly simplified setup involving 3d $\mathcal{N}=2$ theories and $W_{q,t}$ free boson correlators with finitely many screening charge insertions \cite{Aganagic:2013tta,Aganagic:2014oia}. By contrast, in the 5d setup one needs infinitely many of those \cite{Kimura:2015rgi}, and the 3d theories can be understood as codimension 2 defects of the parent 5d theories. In \cite{Nedelin:2016gwu} is was shown that to any $\Gamma$ quiver $W_{q,t}(\Gamma)$ algebra of \cite{Kimura:2015rgi} one can associate a 3d $\mathcal{N}=2$ unitary quiver gauge theory and that there is a dictionary between observables on the two sides. In particular, {\it $\Gamma$ determines the gauge theory quiver}, and the Wilson loop generating function
\be
{\rm Z}=\sum_{\lambda}\langle \mathcal{W}^{{\rm background}}_\lambda \rangle\langle\mathcal{W}^{{\rm gauge}}_\lambda \rangle\nn
\ee
can be identified with a highest weight state of the $W_{q,t}(\Gamma)$ algebra generated by $\{\T^a_n,n\in\mathbb{Z}\}$
\be
{\rm Z}\simeq \ket{\Z}~,\quad \T^a_{n\geq 0}\ket{\Z}\propto \delta_{n,0}\ket{\Z}~,\nn
\ee
where $a$ runs over the nodes of the quiver $\Gamma$. This picture holds true not only when the 3d gauge theory is placed on the flat space $\mathbb{R}^2\times \mathbb{S}^1$, but also on compact spaces $\mathcal{M}_3$ such as $\mathbb{S}^3$, $\mathbb{L}(r,1)= \mathbb{S}^3/\mathbb{Z}_r$ and $\mathbb{S}^2\times \mathbb{S}^1$. In this case the $W_{q,t}(\Gamma)$ algebra is enhanced to its {\it modular double}, meaning that there are two commuting copies of the algebra with ${\rm SL}(2,\mathbb{Z})$-related deformation parameters. This structure nicely reflects the factorization properties of the 3d gauge theory observables \cite{Pasquetti:2011fj,Beem:2012mb}, which are in turn consistent with the Heegaard splitting of $\mathcal{M}_3$ into a pair of solid tori $\mathbb{D}^2\times \mathbb{S}^1\simeq \mathbb{R}^2\times \mathbb{S}^1$ glued by the appropriate ${\rm SL}(2,\mathbb{Z})$ element. 

Since the 3d gauge theory setup can be naturally lifted to 4d by adding a circle direction and trigonometric $W_{q,t}(\Gamma)$ algebras can be further deformed to elliptic $W_{q,t;q'}(\Gamma)$ algebras \cite{Nieri:2015dts,Iqbal:2015fvd,Kimura:2016dys,Tan:2016cky}, we are led to study whether the correspondence described above survives. An affirmative answer was given in \cite{Nieri:2015dts} for the flat space background $\mathbb{R}^2\times\mathbb{T}^2$, and the goal of this paper is to analyze the compact spaces $\mathcal{M}_3\times \mathbb{S}^1$. One of the main motivations for studying the relation between elliptic $W_{q,t;q'}(\Gamma)$ algebras and 4d supersymmetric gauge theories on compact backgrounds is because their partition functions provide important quantities such as the (superconformal) index when $\mathcal{M}_3=\mathbb{S}^3$. Knowing the algebraic structures hidden in the gauge theories may help in having a better understanding of their intricate web of dualities and their relations with integrable models \cite{Bazhanov:2010kz,Bazhanov:2011mz,Spiridonov:2010em,Yamazaki:2012cp,Yamazaki:2013nra,Yagi:2015lha,Maruyoshi:2016caf,Yagi:2017hmj}.

Similarly to the 3d/trigonometric case, we found that on compact spaces the relevant algebra is the {\it elliptic $W_{q,t;q'}(\Gamma)$ modular double}, meaning that there are two commuting copies of the algebra with ${\rm SL}(2,\mathbb{Z})\subset {\rm SL}(3,\mathbb{Z})$-related deformation parameters.\footnote{The notion of elliptic modular double has also appeared in \cite{2008arXiv0801.4137S}, following the construction of \cite{Faddeev:1999fe,Ponsot:1999uf}.} As before, this structure nicely reflects the factorization properties of the 4d gauge theory observables \cite{Yoshida:2014qwa,Peelaers:2014ima,Chen:2014rca,Nieri:2015yia}. There are important differences with respect to the 3d/trigonometric case though. Firstly, we needed to introduce the notion of {\it defect generating function} since supersymmetric Wilson loops cannot be generally defined. This object can be thought of as encoding the v.e.v.'s of BPS surface operators supported on $\mathbb{T}^2\subset \mathcal{M}_3\times \mathbb{S}^1$, and it turns out that the 2d theories of class $\mathcal{H}$ \cite{Gadde:2013sca} play a crucial role. Secondly, while the $W_{q,t,q'}(\Gamma)$ modular double is defined for any quiver $\Gamma$, a satisfactory dual gauge theory description can be given only for a restricted type of quivers due to anomalies. Lagrangian theories of class $\mathcal{S}$ are a relevant subset.

The rest of the paper is organized as follows. In section \ref{sec:gauge} we review the observables of 4d supersymmetric gauge theories which can be computed through localization. In particular, we will introduce the defect generating function which will be the main object of interest from the $W_{q,t;q'}(\Gamma)$ algebra perspective. In section \ref{sec:ellW} we review basic facts about $W_{q,t;q'}(\Gamma)$ algebras.
In section  \ref{sec:ellmodW} we introduce the $W_{q,t;q'}(\Gamma)$ modular double and the associated matrix models. In section \ref{sec:gaugeW} we establish the relation between gauge theories and $W_{q,t;q'}(\Gamma)$ algebras. In section \ref{sec:concl} we discuss our results further, with particular emphasis on open questions and future developments. In appendix \ref{sec:specialf} we have recalled the definition and properties of the special functions appearing in this paper, and in appendix \ref{sec:torus} we have briefly discussed the relation between trigonometric and elliptic algebras. 

\section{Gauge theories on supersymmetric backgrounds}\label{sec:gauge}
Following the works \cite{Festuccia:2011ws,Dumitrescu:2012ha,Klare:2012gn,Klare:2013dka}, 4d $\mathcal{N}=1$ gauge theories with an R-symmetry can be put on curved spaces while preserving some supersymmetry. If we require to preserve at least two supercharges of opposite R-charge, the allowed manifolds are $\mathbb{T}^2$ fibrations over a Riemann surface. Focusing on the genus zero case, in the following subsections we will review some aspect of these theories, in particular the computation of supersymmetric observables through the matrix models obtained from localization (for a comprehensive review  see \cite{Pestun:2016jze}). 

\subsection{$\mathbb{S}^3\times \mathbb{S}^1$}
A particularly relevant example of supersymmetry preserving compact 4-manifolds is provided by Hopf surfaces, which are diffeomorphic to $\mathbb{S}^3\times \mathbb{S}^1$. They have an interesting 2-parameter family of complex structures with moduli $\p,\q\in\mathbb{C}^{\times}$, $0<|\q|\leq |\p|<1$, and the geometry we are interested in can be defined by the quotient 
\be
\mathbb{C}^2\backslash \{(0,0)\}\ni(z_1,z_2)\sim (\q z_1,\p z_2)~.
\ee
The partition functions of Lagrangian $\mathcal{N}=1$ gauge theories with gauge group $G$\footnote{In this paper we will be only interested in ${\rm U}(N)$ or ${\rm SU}(N)$ gauge groups.}  can be computed by localization \cite{Assel:2014paa}, yielding the matrix model\footnote{For a collection of variables $\ul x=\{x_k,k=1,\ldots,N\}$, we will use the shorthand notation $\frac{\rd \ul x}{2\pi\i\ul x}=\prod_{k=1}^N\frac{\rd x_k}{2\pi\i x_k}$.}$^{,}$\footnote{We will always omit constants such as $(\q;\q)_\infty^{|G|}(\p;\p)_\infty^{|G|}/|{\rm Wey}|$.}
\be\label{Z:S3xS1}
{\rm Z}[\mathbb{S}^3\times \mathbb{S}^1]=\e^{-\i\pi \mathcal{P}_3(\ln\ul \zeta)}\oint_{\mathbb{T}^{|G|}}\frac{\rd \ul z}{2\pi\i\ul z}\;\Delta_{1\,\rm vec}(\ul z)\Delta_{\rm chi}(\ul z,\ul \zeta)~,
\ee
where 
\be
\Delta_{1\,\rm vec}(\ul z)=\prod_{\alpha\neq 0}\frac{1}{\Gamma(\bs z^\alpha;\p,\q)}~,\quad \Delta_{\rm chi}(\ul z,\ul \zeta)=\prod_I\prod_{\rho,\phi\in\mathcal{R}_I}\Gamma((\p\q)^{R_I/2}\bs z^\rho\bs \zeta^\phi;\p,\q)
\ee
are the 1-loop contributions of vector multiplets\footnote{We are including the Haar measure and gauge fixing contributions.} and chirals in representations $\mathcal{R}_I$ of the gauge and flavor groups, $\bs z$ and $\bs \zeta$ are gauge and global holonomies in the Cartan tori, $\alpha$ is a root of the Lie algebra of $G$ and $\rho,\phi$ are weights of $\mathcal{R}_I$. The overall exponent $\mathcal{P}_3$ is a cubic polynomial in the logarithm of the global fugacities. For every ${\rm U}(1)$ factor in the gauge group we may also consider a Fayet-Iliopoulos (FI) parameter $\kappa$ (discretized) contributing to the integrand with
\be
z_{{\rm U}(1)}^\kappa~.
\ee
The $\mathbb{S}^3\times \mathbb{S}^1$ partition function can also be interpreted as the $\mathcal{N}=1$ index \cite{Romelsberger:2005eg,Kinney:2005ej}
\be
{\rm Z}[\mathbb{S}^3\times \mathbb{S}^1]\propto {\rm I}_{1}=\textrm{Tr}(-1)^{\bs F} \q^{\bs H_1}\p^{\bs H_2}(\p \q)^{\frac{\bs R}{2}}\bs \zeta~,
\ee
where $\bs F$ is the fermion number, $\bs H_{1,2}$ are the generators for rotations in the $z_{1,2}$ planes, $\bs R$ is the R-symmetry and the trace is taken over  the Hilbert space of the theory on $\mathbb{S}^3$. By construction, only the states annihilated by a selected supercharge $\mathcal{Q}$ contribute to the index, namely those satisfying $\{\mathcal{Q},\mathcal{Q}^\dagger\}=E-H_1-H_2-3R/2=0$.

There are two specially interesting situations. The first one is when each vector is accompanied by an adjoint chiral, so that the multiplet content is that of the $\mathcal{N}=2$ vector. They contribute  to the integral representation of the index with the 1-loop factor
\be\label{Delta2}
\Delta_{2\,\rm vec}(\ul z)=\prod_{\alpha\neq 0}\frac{\Gamma(\hat\t\bs z^\alpha;\p,\q)}{\Gamma(\bs z^\alpha;\p,\q)}~,
\ee
where for later convenience we defined
\be
\hat\t=(\p\q)^{\frac{R_{\rm ad}}{2}}~.
\ee
If the theory is effectively $\mathcal{N}=2$ then the R-symmetry is at least ${\rm SU}(2)_{R_2}$, but we consider the case when it is ${\rm SU}(2)_{R_2}\times {\rm U}(1)_{r_1}$ as for superconformal theories. Then one can turn on another fugacity $\t$ and define the $\mathcal{N}=2$ (superconformal) index
\be
{\rm I}_{2}=\textrm{Tr}(-1)^{\bs F} \q^{\bs H_1}\p^{\bs H_2}(\p \q\t^{-1})^{-\bs r_1}\t^{\bs R_{2}}\bs \zeta~,
\ee
where the only contributing states are those satisfying $\{\mathcal{Q},\mathcal{Q}^\dagger\}=E-H_1-H_2-2 R_2+r_1=0$.\footnote{From the $\mathcal{N}=1$ perspective, $R=\frac{2}{3}(2R_2-r_1)$.} The measure (\ref{Delta2}) can be interpreted from this perspective, in which case $\hat\t=\p\q\t^{-1}$. Of course, the chiral multiplets must come in pairs forming hypers, which contribute to the integrand of the index with
\be
\Delta_{\rm hyp}(\ul z,\ul \zeta)=\prod_I\prod_{\rho,\phi\in\mathcal{R}_I}\frac{\Gamma((\p\q)^{1/2}\hat\t^{-1/2}\bs z^\rho\bs \zeta^\phi;\p,\q)}{\Gamma((\p\q)^{1/2}\hat\t^{1/2}\bs z^\rho\bs \zeta^\phi;\p,\q)}~.
\ee

The second special situation is when the $\mathcal{N}=2$ vector is accompanied by an adjoint hyper hence forming the multiplet content of the $\mathcal{N}=4$ vector. If the theory is effectively $\mathcal{N}=4$, the R-symmetry group is ${\rm SU}(4)$ with Cartan generators $\bs r_{1,2,3}$ and one can turn on another fugacity $\mathfrak{m}$ associated with the $\mathcal{N}=2^*$ deformation and define the $\mathcal{N}=4$ index
\be
{\rm I}_{4}=\textrm{Tr}(-1)^{\bs F} \q^{\bs H_1}\p^{\bs H_2}(\p \q)^{\frac{\bs r_1-\bs r_3}{2}}\hat\t^{\bs r_2+\bs r_3}\mathfrak{m}^{\bs r_3}~,
\ee
which receives contributions only from the states satisfying $\{\mathcal{Q},\mathcal{Q}^\dagger\}=E-H_1-H_2 -3r_1/2-r_2-r_3/2=0$. The integrand of index is then
\be
\Delta_{4\,\rm vec}(\ul z)=\prod_{\alpha\neq 0}\frac{\Gamma(\hat\t \bs z^\alpha;\p,\q)}{\Gamma(\bs z^\alpha;\p,\q)}\frac{\Gamma(\mathfrak{m}\bs z^\alpha;\p,\q)}{\Gamma(\hat\t\mathfrak{m} \bs z^\alpha;\p,\q)}~.
\ee
From the $\mathcal{N}=1$ viewpoint we can treat $\hat{\mathfrak t},\mathfrak{m}$ as additional parameters. The vector contributions $\Delta_{1,2,4\,\rm vec}(\ul z)$ are part of the measures that we will analyze in the second half of the paper from the elliptic $W_{q,t;q'}(\Gamma)$ algebra perspective.

\subsection{Defect generating function}
The partition function is the simplest supersymmetric observable and the exact computation of more sophisticated quantities is desirable. The supersymmetric gauge theories that we are considering can be enriched by the inclusion of different types of BPS surface defects supported on a torus \cite{Gadde:2013ftv}. The torus arises from the $\mathbb{S}^1$ circle and the Hopf fiber of the $\mathbb{S}^3$, and it can be identified with
\be
\mathbb{T}^2\simeq \{z_1\in\mathbb{C}^\times~|~z_1\sim \q z_1\}~.
\ee
Of course, we also have the other choice $z_2\sim \p z_2$, the difference being the location of defect at the North or South pole of the Hopf base. The surface operators we are interested in can be more easily defined in the UV by coupling 2d degrees of freedom to the bulk. This can be done by considering a 2d theory with a global symmetry which is a subgroup of the bulk gauge group and gauging it. Since we will be interested in computing partition functions of 4d theories with defects, we must supplement the integrand of the 4d partition functions with additional 1-loop factors encoding the contribution of the 2d degrees of freedom. 

The torus partition function of a 2d theory with $\mathcal{N}=(2,2)$ supersymmetry defines the elliptic genus 
\be
{\rm I}_{(2,2)}=\textrm{Tr}(-1)^{\bs F} \q^{\bs H_L}\mathfrak{y}^{\bs J_L}\bs \zeta~,
\ee
where $\bs H_{(L)R}$, $\bs J_{(L)R}$ are the (left) right-moving Hamiltonian and R-symmetry respectively, while  $\bs \zeta$ is an element of the global Cartan torus. The only states contributing are those satisfying $\{\mathcal{Q},\mathcal{Q}^\dagger\}=2H_R-J_R=0$. In order to couple the 2d degrees of freedom to the bulk, one has to embed the 2d algebra into the 4d one, which can be achieved through the identifications \cite{Gadde:2013ftv}
\be
\bs H_1=\bs H_L-\bs H_R~,\quad 4\bs R_2+2\bs H_2=\bs J_L+\bs J_R~,\quad 2r_1=\bs J_L-\bs J_R~, 
\ee
implying
\be
\mathfrak{y}=\q^\frac{1}{2}\hat\t^{-1}~.
\ee
From the 2d perspective there is a global ${\rm U}(1)$ generated by $\bs H_2+\bs R_2$ and coupling to $\zeta=\hat{\mathfrak{t}}\p \q^{-1}$. If the 2d theory has only $\mathcal{N}=(0,2)$ supersymmetry the elliptic genus is
\be
{\rm I}_{(0,2)}=\textrm{Tr}(-1)^{\bs F} \q^{\bs H_L}\bs \zeta~.
\ee
The embedding of the 2d algebra is achieved through the identifications
\be
\bs H_1=\bs H_L-\bs H_R~,\quad \frac{3}{2}\bs R+\bs H_2=\bs J_R~,
\ee
and now there is a global ${\rm U}(1)$ generated by $\bs H_2+\bs R/2$ and coupling to $\zeta=\p \q^{-1/2}$. In $\mathcal{N}=(0,2)$ language, the basic multiplets of a gauge theory with gauge group $G^{2\rm d}$ and matter in representations $\mathcal{R}^{2\rm d}_I$ of the 2d gauge and global symmetry groups are vector, chiral and Fermi multiplets, whose elliptic genera are \cite{Benini:2013nda,Benini:2013xpa}
\be
\begin{split}
Z^{\rm 2d}_{\rm vec}(\ul z)&=\prod_{\alpha\neq 0}\Theta(\bs z^\alpha;\q)~,\\
Z^{\rm 2d}_{\rm chi}(\ul z,\ul\zeta)&=\prod_I\prod_{\rho,\phi\in\mathcal{R}^{2\rm d}_I}\frac{1}{\Theta(\bs z^\rho \q^{H_{L,I}}\bs\zeta^\phi;\q)}~,\quad Z^{\rm 2d}_{\rm F}(\ul z,\ul\zeta)=\prod_I\prod_{\rho,\phi\in\mathcal{R}^{2\rm d}_I}\Theta(\bs z^\rho \q^{H_{L,I}}\bs\zeta^\phi;\q)~,
\end{split}
\ee
where $\bs z$ and $\bs \zeta$ are in the maximal torus of the gauge and global symmetry groups, $\alpha$ is a root of the Lie algebra of $G^{\rm 2d}$ and $\rho,\phi$ are weights of $\mathcal{R}^{\rm 2d}_I$. Given these ingredients we can now easily couple a defect to the bulk at the North (N) and/or South (S) poles.

Let us now suppose that the 4d gauge group is $G=\bigtimes_{a} {\rm U}(N_a)$.\footnote{We can consider ${\rm SU}(N_a)$ groups by imposing a $\delta$-function constraint on the ${\rm U}(N_a)$ gauge variables. ${\rm SU}(N_a)$ groups may actually be needed to avoid anomalies, we will discuss this point later on in section \ref{sec:gaugeW}.} Instead of coupling a specific defect, it is convenient to define the {\it defect generating function} which depends on a set of {\it Miwa} or {\it time variables} $\ul\tau^a_{i=\rm N,S}$, that is
\be\label{DefGenTau}
\boxed{
{\rm Z}[\mathbb{S}^3\times \mathbb{S}^1](\{\tau_{0,i}^a,\ul\tau^a_i\})=\e^{\mathcal{N}_0(\tau_{0,i})}\oint_{\mathbb{T}^{|G|}}\frac{\rd \ul z}{2\pi\i\ul z}\;\Delta^{\rm 4d}(\ul z) \prod_a \exp\left(-\sum_{j=1}^{N_a}\sum_{n\neq 0}\frac{\tau^a_{n,i}z_{j,a}^n}{n(1-\q_i^n)}\right)}~,
\ee
where $\e^{\mathcal{N}_0(\tau_{0,i})}$ is a normalization and $\Delta^{\rm 4d}(\ul z)$ denotes one of the 4d measures we have discussed before. For later convenience we have also defined
\be
\q_{i=\rm N}=\q~,\quad \q_{i=\rm S}=\p~.
\ee
The reason is that the defect generating function is a quite general object since it encodes {\it all the observables} which can be computed through Coulomb branch localization, and this is the kind of object that we will study from the elliptic $W_{q,t;q'}$ algebra perspective in later sections. In order to substantiate this claim we give two relevant examples. Firstly, the inclusion of a 4d ${\rm U}(N_a)$ (anti-) fundamental chiral multiplet with global ${\rm U}(1)$ fugacity $\zeta$ simply amounts to shifting the time variables by the following background
\be
\square:~\tau^a_{n,{\rm N,S}}\to \tau^a_{n,{\rm N,S}}-\frac{\zeta^n}{1-\q_{\rm S,N}^n}~,\quad \bar\square:~\tau^a_{n,{\rm N,S}}\to \tau^a_{n,{\rm N,S}}+\frac{\p^n\q^n \zeta^{-n}}{1-\q^n_{\rm S,N}}~,\quad n\neq 0~.
\ee
The appearance of the opposite rotational fugacity $\q_{\rm S,N}$ in the denominator means that we are correctly considering bulk degrees of freedom. Secondly, we can couple 2d \mbox{(anti-)} fundamental chiral and/or Fermi multiplets with global ${\rm U}(1)$ fugacity $\zeta$ at the North and/or South poles by considering the backgrounds
\be
\square:~\tau^a_{n,i}\to \tau^a_{n,i}\mp \zeta^n~,\quad \bar\square:~\tau^a_{n,i}\to \tau^a_{n,i}\mp \q_i^n\zeta^{-n}~,\quad n\neq 0~.
\ee
The coupling of more complicated defects can be achieved by acting with suitable operators on the generating function. See also \cite{Gaiotto:2012xa,Gadde:2013ftv,Bullimore:2014nla} for a closely related discussion. 

It may seem that the generating function is only a useful tool without a clear gauge theory interpretation. Actually, we can try to relate it to the generating function (and hence the origin of its name) of v.e.v.'s of a wide class (complete in the sense to be clarified below) of defect theories supported on $\mathbb{T}^2$. The idea is to find a suitable family of BPS defect operators $\mathcal{D}^i_\lambda$ characterized by a discrete label $\lambda$ and a global symmetry group such that their values in a given supersymmetric background 
\be
\langle\mathcal{D}^i_\lambda\rangle_{\ul\zeta,\ul v}=d_\lambda(\ul\zeta,\ul v;\q_i)
\ee
are complete
\be
 \sum_{\lambda}d_\lambda(\ul\zeta,\ul v;\q_i)^\vee d_\lambda(\ul\phi,\ul u;\q_i)=\prod_{j,k}\frac{\Theta(\phi_j\zeta_k;\q_i)}{\Theta(\phi_j u_k;\q_i)\Theta(\zeta_k v_j;\q_i)}~,
\ee 
where $\ul\zeta,\ul v,\ul\phi,\ul u$ are fugacities for global symmetries and $^\vee$ is some dual operation. If we now interpret the r.h.s. as the 1-loop determinant of a bunch of free chiral and Fermi multiplets, we can couple them to the ${\rm U}(N_a)$ node of the 4d bulk theory by gauging $\ul\phi$ for instance, assuming $j=1,\ldots, N_a$. The partition function of the resulting 4d-2d coupled system is
\be\label{DefGenZeta}
{\rm Z}[\mathbb{S}^3\times \mathbb{S}^1](\ul\zeta,\ul v;\ul u)=\frac{1}{\prod_{j,k}\Theta(\zeta_k v_j;\q_i)}\oint_{\mathbb{T}^{|G|}}\frac{\rd \ul z}{2\pi\i\ul z}\;\Delta^{\rm 4d}(\ul z)\prod_{j=1}^{N_a} \prod_{k}\frac{\Theta(\zeta_{k} z_{j,a} ;\q_i)}{\Theta(u_k z_{j,a};\q_i)}~.
\ee
Using the expansion
\be
\prod_k \Theta(x_k;\q_i)=\exp\left(-\sum_{n\neq 0}\frac{\sum_k x_k^n}{n(1-\q_i^n)}\right)~,
\ee
we can introduce Miwa variables for $\ul\zeta,\ul u$
\be
\tau^a_{n,i}=\sum_k \zeta^n_k~,\quad \eta^a_{n,i}=\sum_k u^n_k~,
\ee
and find that
\begin{multline}
\e^{\mathcal{N}_0(\tau_{0,i})}\;\prod_{j,k}\Theta(\zeta_k v_j;\q_i)\;{\rm Z}[\mathbb{S}^3\times \mathbb{S}^1](\ul\zeta,\ul v;\ul u)=\\
=\e^{\mathcal{N}_0(\tau_{0,i})}\;\prod_{j,k}\Theta(\zeta_k v_j;\q_i)
\sum_\lambda d_\lambda(\ul\zeta,\ul v;\q_i)^\vee\!\! \oint_{\mathbb{T}^{|G|}}\frac{\rd \ul z}{2\pi\i\ul z}\Delta^{\rm 4d}(\ul z)\; d_\lambda(\ul z_a , \ul u;\q_i)=\\
={\rm Z}[\mathbb{S}^3\times \mathbb{S}^1](\{\tau_{0,i}^a,\ul\tau^a_i-\ul\eta^a_i\})~,
\end{multline}
coinciding with the generating function (\ref{DefGenTau}) with a background $-\ul\eta_i^a$ for the time variables. Finally, assuming that the defect v.e.v.'s are orthogonal w.r.t. some integral measure
\be\label{Dmeasure}
\oint [\rd \ul \phi]\; d_{\lambda'}(\ul\phi,\ul u)\; d_\lambda(\ul\phi,\ul u)\propto \delta_{\lambda,\lambda'}~,\quad \oint [\rd \ul v]\; d_{\lambda'}(\ul\zeta,\ul v)^\vee\; d_\lambda(\ul\zeta,\ul v)^\vee\propto \delta_{\lambda,\lambda'}~,
\ee
one can in principle extract the v.e.v. of $\mathcal{D}^i_\lambda$ in the 4d gauge theory by projection
\be
\langle\mathcal{D}^i_\lambda\rangle_{\ul u}=\oint_{\mathbb{T}^{|G|}}\frac{\rd \ul z}{2\pi\i\ul z}\;\Delta^{\rm 4d}(\ul z)\; d_\lambda(\ul z_a , \ul u;\q_i) \propto \oint [\rd \ul v]\;  d_{\lambda}(\ul\zeta,\ul v)^\vee\; Z[\mathbb{S}^3\times \mathbb{S}^1](\ul\zeta,\ul v;\ul u)~.
\ee
However, this formal reasoning must be finalized by the actual definition of the observables $\mathcal{D}^i_\lambda$, if there is any. In order to get an intuition about the solution to this problem, it is useful to consider first the simpler setup given by the dimensionally reduced 3d theory on the squashed $\mathbb{S}^3$ \cite{Hama:2011ea,Imamura:2011wg,Alday:2013lba,Hosomichi:2014hja}. In this case we can consider supersymmetric Wilson loop operators $\mathcal{W}^i_\lambda$ wrapping the Hopf fiber at North and/or South poles of the base. In a given supersymmetric background, the operators are given by Schur polynomials or characters of the relevant unitary group \cite{Kapustin:2009kz,Tanaka:2012nr}
\be
\langle \mathcal{W}^i_\lambda \rangle_{\ul\zeta} = s_\lambda (\ul \zeta)~,
\ee
where $ \bs \zeta=\e^{\frac{2\pi\i}{\omega_i}\bs \Xi}$ is a fugacity and $\bs \Xi$ is a constant vector multiplet scalar in the Cartan algebra, $\lambda$ is a Young diagram associated to an arbitrary representation and $\omega_{i=1,2}$ are the (imaginary) squashing parameters of the $\mathbb{S}^3$, which are related to the $\q,\p$ parameters according to
\be\label{S3pq}
\q_{i=\rm N}=\q=\e^{2\pi\i\frac{\omega_1}{\omega_3}}~,\quad \q_{i=\rm S}=\p=\e^{2\pi\i\frac{\omega_2}{\omega_3}}~,
\ee
$\omega_3$ being the inverse $\mathbb{S}^1$ radius. As is well known, these line operators/characters satisfy completeness and orthogonality relations, and the Wilson loop generating function was the main object studied in \cite{Nedelin:2016gwu}. Since we are looking for the $\mathbb{S}^1$ lift of this picture, in the 4d setup it is natural to consider the insertion of (normalized) {\it affine characters} into the $\mathbb{S}^3\times \mathbb{S}^1$ partition function. Now the problem is to understand how and if these objects are related to torus defects. Remarkably, the 2d defect theories we are interested in do indeed exist and they were constructed in \cite{Gadde:2013sca}. These theories are called class $\mathcal{H}$ and can be defined by (twisted) compactifications of the 6d $(2,0)$ theory on ``internal" 4-manifolds $\mathcal{M}^{\rm int}_4$, hence denoted by $\mathcal{T}[\mathcal{M}^{\rm int}_4]$. When $\mathcal{M}^{\rm int}_4$ is obtained from the resolution of an ADE singularity, this construction allows one to define a family of 2d $\mathcal{N}=(0,2)$ theories labeled by the corresponding Lie algebra $\mathfrak{g}$ and an integrable representation $\lambda$ of the affine Lie algebra $\hat{\mathfrak{g}}$ at level $M$.\footnote{More generally, the Dynkin diagram should be replaced by the plumbing graph of $\mathcal{M}_4^{\rm int}$.} Here we are assuming that the ``internal" 4d gauge group is ${\rm U}(M)$, and we are interested in the simplest cases $\mathfrak{g}\in\{\mathfrak{u},\mathfrak{su}\}$. The elliptic genus of a class $\mathcal{H}$ theory computes the Vafa-Witten partition function \cite{Vafa:1994tf} of the associated 4-manifold and equals the affine character
\be
{\rm I}_{(0,2)}[\mathcal{T}[\mathfrak{g},\lambda,{\rm U}(M)]]=\chi_\lambda^{\hat{\mathfrak{g}}_M}(\ul \zeta;\q_i)~.
\ee
Here the global 2d fugacity $\bs \zeta$ appears as an element of the maximal torus of $\mathfrak{g}$. The affine characters satisfy the completeness and orthogonality relations  \cite{Gadde:2013sca}
\begin{align}
&\sum_\lambda \chi_{\lambda^\vee}^{\hat{\mathfrak{su}}(M)_{N}}(\ul\zeta;\q_i)\chi_{\lambda}^{\hat{\mathfrak{u}}(N)_{M}}(\ul \phi;\q_i) =\q_i^{-\frac{N M}{24}}\prod_{j=1}^{N}\prod_{k=1}^M\Theta(-\q_i^{1/2}  \zeta_k \phi_j;\q_i)~,\\
&\oint\frac{\rd^N \ul\phi}{2\pi\i\ul\phi}\prod_{j\neq \ell}\Theta(\phi_\ell/\phi_j;\q_i)\;\chi_{\lambda}^{\hat{\mathfrak{u}}(N)_{M}}(\ul \phi;\q_i)\;\chi_{\lambda'}^{\hat{\mathfrak{u}}(N)_{M}}(\ul \phi;\q_i)\propto \delta_{\lambda,\lambda'}~,
\end{align}
from which one can deduce that the observables we are interested in are given by
\be
\begin{split}
d_\lambda(z_a ,\ul u;\q_i) &=\frac{\chi_{\lambda}^{\hat{\mathfrak{u}}(N_a)_{M}}(\ul z_a;\q_i)}{\q_i^{-N_a M/48}\prod_{j=1}^{N_a}\prod_{k=1}^M \Theta(z_{j,a}u_k;\q_i)} ~,\\
d_\lambda(\ul \zeta , \ul v;\q_i)^\vee &=\frac{\chi_{\lambda^\vee}^{\hat{\mathfrak{su}}(M)_{N_a}}(\ul \zeta;\q_i)}{\q_i^{-N_a M/48}\prod_{j=1}^{N_a}\prod_{k=1}^M \Theta(\zeta_k v_j;\q_i)} ~,
\end{split}
\ee
and that the dual operation $^\vee$ corresponds to exchanging level/rank $\mathfrak{u}(N_a)_M\to \mathfrak{su}(M)_{N_a}$ and transposing the diagram $\lambda\to\lambda^\vee$. Also, the measure (\ref{Dmeasure}) becomes 
\be
[\rd \ul \phi]=\frac{\rd^N \ul\phi}{2\pi\i\ul\phi}\prod_{j\neq \ell}\Theta(\phi_\ell/\phi_j;\q_i)\prod_{j=1}^N\prod_{k=1}^M \Theta(\phi_j u_k;\q_i)^2~.
\ee


\subsection{Other spaces and factorization}\label{sec:d2xt2}
As we have already mentioned, 4d $\mathcal{N}=1$ theories can be defined on other spaces in addition to $\mathbb{S}^3\times \mathbb{S}^1$. In this section we will briefly extend the previous discussion to $\mathbb{L}(r,1)\times \mathbb{S}^1$ and $\mathbb{S}^2\times \mathbb{T}^2$. Partition functions on these spaces were computed in \cite{Benini:2011nc,Razamat:2013jxa,Nishioka:2014zpa,Closset:2013sxa,Honda:2015yha}, and for unitary gauge groups they take a form similar to (\ref{Z:S3xS1})
\begin{align}
{\rm Z}[\mathbb{L}(r,1)\times \mathbb{S}^1]&=\sum_{\ul \ell \in\mathbb{Z}_r^{|G|}}\oint_{\mathbb{T}^{|G|}}\frac{\rd \ul z}{2\pi\i\ul z}\;\Delta_{\mathbb{L}(r,1)\times \mathbb{S}^1}(\ul z,\ul\ell,\ul \zeta,\ul h)~,\\
{\rm Z}[\mathbb{S}^2\times \mathbb{T}^2]&=\sum_{\ul \ell \in\mathbb{Z}^{|G|}}\oint_{\rm J.K.}\frac{\rd \ul z}{2\pi\i\ul z}\;\Delta_{\mathbb{S}^2\times\mathbb{T}^2}(\ul z,\ul\ell,\ul \zeta,\ul h)~,
\end{align}
where we have included the zero-point energies in the definition of the total measures and $\rm J.K.$ denotes the Jefferey-Kirwan residue prescription. As before, the continuous variables $\bs z=\e^{2\pi\i \bs X}$ and $\bs \zeta=\e^{2\pi\i\bs \Xi}$ denote the gauge and background holonomies along $\mathbb{S}^1$ or $\mathbb{T}^2$, while the discrete variables $\bs \ell$ and $\bs h$ parametrize gauge and background holonomies along the non-contractible circle of $\mathbb{L}(r,1)$ or fluxes through $\mathbb{S}^2$. The partition function on $\mathbb{L}(r,1)\times \mathbb{S}^1$ is also proportional to the so-called lens index, and at $r=1$ one recovers the $\mathbb{S}^3\times \mathbb{S}^1$ geometry and the results discussed in the previous subsection. A peculiarity of the $\mathbb{S}^2\times \mathbb{T}^2$ background is that there is a unit flux for the R-symmetry and hence the R-charges must be integers. The 3d limit of the $\mathbb{S}^2\times \mathbb{T}^2$ partition function coincides with the A-twisted index \cite{Benini:2015noa}, while the traditional 3d index \cite{Imamura:2011su,Kapustin:2011jm} can be obtained from the $r\to\infty$ limit of the lens index. 

The partition functions associated to  the different compact spaces $\mathcal{M}_3\times \mathbb{S}^1$, with $\mathcal{M}_3\in\{\mathbb{S}^3,\mathbb{L}(r,1),\mathbb{S}^2\times \mathbb{S}^1\}$, look very different. However,  they can be recovered by a gluing prescription of 1-loop determinants of the theory defined on the half-space $\mathbb{D}^2\times \mathbb{T}^2$ \cite{Nieri:2015yia}, which are given by\begin{figure}
\begin{center}
\includegraphics[height=0.18\textheight]{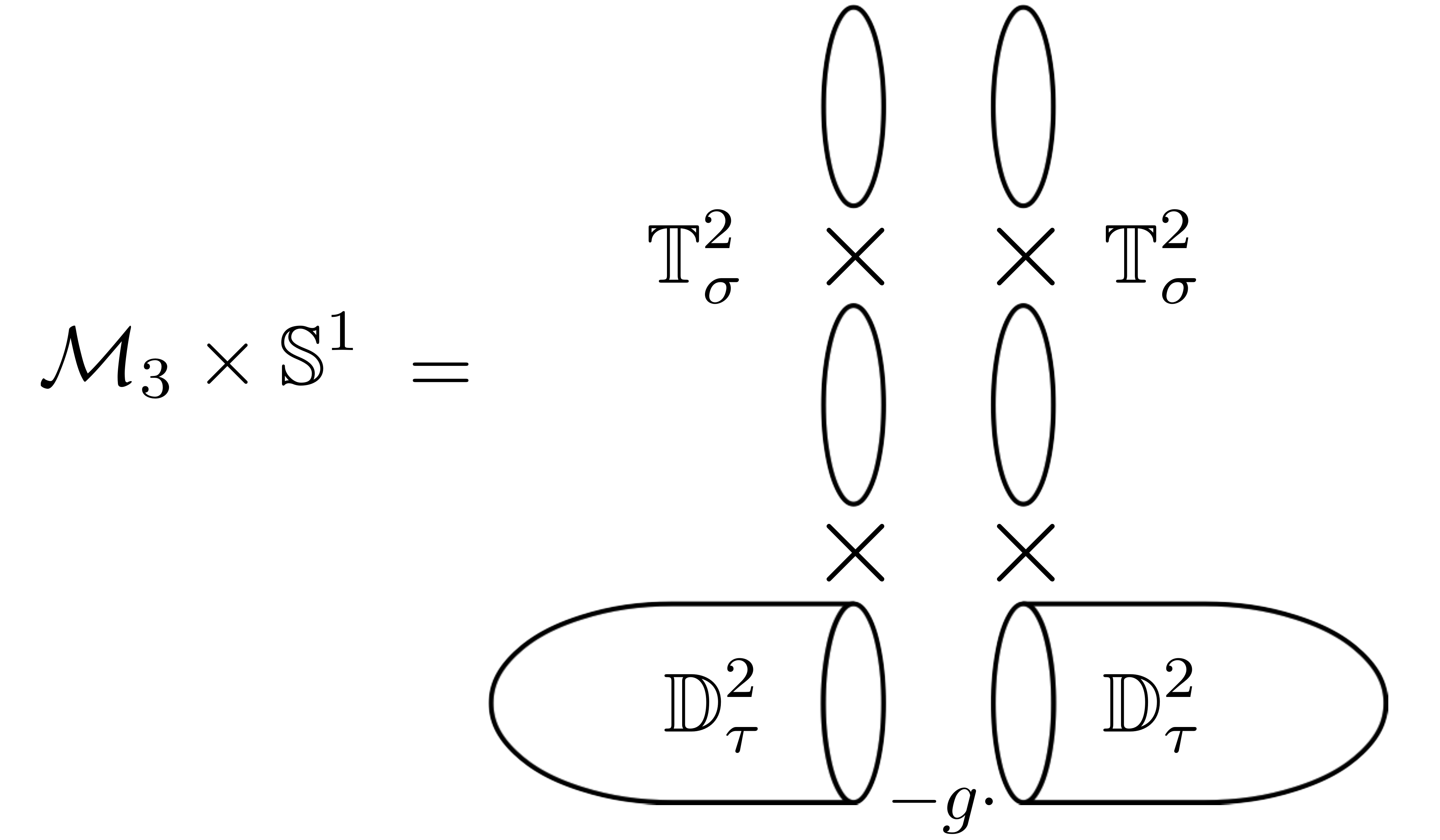}
\end{center}
\caption{Decomposition of $\mathcal{M}_3\times \mathbb{S}^1$. The $g\in {\rm SL}(2,\mathbb{Z})\subset{\rm SL}(3,\mathbb{Z})$ action is on all the geometric and bundle moduli.}
\label{m3s1}
\end{figure}
\be\label{Z1loop:d2xt2}
\begin{split}
\Upsilon_{\rm vec}(\ul w)&=\prod_{\alpha\neq 0}\frac{1}{\Gamma(\bs w^\alpha;q_\tau,q_\sigma)}~,\\
\Upsilon_{\rm chi}^{\rm N}(\ul w)&=\prod_I\prod_{\rho,\phi\in\mathcal{R}_I}\Gamma(\bs w^{\rho} \bs \xi^{\phi};q_\tau,q_\sigma)~,\quad \Upsilon_{\rm chi}^{\rm D}(\ul w)=\prod_I\prod_{\rho,\phi\in\mathcal{R}_I}\frac{1}{\Gamma(q_\tau \bs w^{-\rho} \bs \xi^{-\phi};q_\tau,q_\sigma)}~,
\end{split}
\ee
where $\rm N$ and $\rm D$ denote Neumann or Dirichlet type boundary conditions and $\bs w,\bs \xi$ are elements of the maximal torus of the gauge and global symmetry groups respectively. As before, $\alpha$ is a root of the Lie algebra of the gauge group and $\rho,\phi$ are weights of the matter representations $\mathcal{R}_I$. The fugacities 
\be
q_\tau=\e^{2\pi\i\tau}~,\quad q_\sigma=\e^{2\pi\i\sigma}
\ee
encode the fibration moduli, namely $\tau$ is the disk equivariant parameter\footnote{We hope there will be no confusion between the equivariant parameter $\tau$ and the time variables $\tau^a_n$.} and $\sigma$ the torus modular parameter, which can be in turn identified with the moduli of the boundary $\mathbb{T}^3$. Given the total 1-loop determinant $\Upsilon(\ul w;\tau,\sigma)$ of the theory on the half-space, the compact space integrands are given by
\be\label{Z:PUU}
\boxed{\Delta_{\mathcal{M}_3\times \mathbb{S}^1}(\ul z,\ul \ell,\ul \zeta,\ul h)=\e^{-\i\pi\mathcal{P}_3(\ul X,\ul \Xi)}\;\Upsilon(\ul w;\tau,\sigma)^{(-g)}\;\Upsilon(\ul w;\tau,\sigma)}~,
\ee
where $^{(-g)}$ is an involution acting on the half-space variables $\bs w,\bs \xi,\tau,\sigma$ as the appropriate $g\in {\rm SL}(2,\mathbb{Z})\subset{\rm SL}(3,\mathbb{Z})$ element (composed with the inversion, see e.g. figure \ref{m3s1}). The exponent $\mathcal{P}_3(\ul X,\ul \Xi)$ is a cubic polynomial of its arguments, and it turns out to encode the various gauge, mixed-gauge and global anomalies. On geometrical grounds, the $g$ element should be
\be
\arraycolsep=1.4pt\def\arraystretch{1}
\begin{array}{c|c|c|c}
~&~\mathbb{S}^3\times \mathbb{S}^1~&~\mathbb{L}(r,1)\times \mathbb{S}^1~&~\mathbb{S}^2\times\mathbb{T}^2\\
\hline
~g~&~S=\left(\begin{array}{cc}0&-1\\1&0\end{array}\right)~&~ST^rS=\left(\begin{array}{cc}-1&0\\r&-1\end{array}\right)~&~id=\left(\begin{array}{cc}1&0\\0&1\end{array}\right)
\end{array}~~~~~,
\ee
and in fact the $g$ action can be summarized in the following tables:
\be\label{ggluing}
\begin{split}
\begin{array}{c|c|c||c|c}
~&\bs w&\bs \xi&\bs w^{(-g)}&\bs\xi^{(-g)}\\
\hline
\mathbb{L}(r,1)\times \mathbb{S}^1&\e^{2\pi\i\bs X}\e^{\frac{2\pi\i}{r}\bs\ell}&\e^{2\pi\i\bs \Xi}\e^{\frac{2\pi\i}{r}\bs h}&\e^{-2\pi\i g\cdot \bs X}\e^{-\frac{2\pi\i}{r}g\cdot\bs\ell}&\e^{-2\pi\i g\cdot \bs \Xi}\e^{-\frac{2\pi\i}{r}g\cdot\bs h}\\
\hline
\mathbb{S}^2\times\mathbb{T}^2&\e^{2\pi\i\bs X}\e^{-\i\pi\tau\bs\ell}&\e^{2\pi\i \bs \Xi}\e^{-\i\pi\tau\bs\ell}&\e^{-2\pi\i g\cdot\bs X}\e^{-\i\pi g\cdot\tau\bs\ell}&\e^{-2\pi\i \bs g\cdot \Xi}\e^{-\i\pi g\cdot \tau \bs\ell}
\end{array}\\
\begin{array}{c|c|c|c|c|c|c}
~&-g\cdot \tau&-g\cdot \sigma&-g\cdot \bs X&-g\cdot \bs \Xi&-g\cdot \bs \ell &-g\cdot \bs h\\
\hline
\mathbb{L}(r,1)\times \mathbb{S}^1&\frac{\tau}{r\tau-1}&\frac{\sigma}{r\tau-1}&\frac{\bs X}{r\tau-1}&\frac{\bs \Xi}{r\tau-1}&r-\bs\ell&r-\bs h\\
\hline
\mathbb{S}^2\times\mathbb{T}^2&-\tau&-\sigma&\bs X&\bs \Xi&\bs\ell&\bs h
\end{array}~~~~~~~~~~~.
\end{split}
\ee
%

Notice that we have not explicitly included $\mathbb{S}^3\times \mathbb{S}^1$ in the list because for our purposes it will be more convenient to regard $\mathbb{S}^3$ as  $\mathbb{L}(1,1)$. This is not in contradiction with the statement that the $\mathbb{S}^3\times \mathbb{S}^1$ 1-loop integrand can be obtained by fusing two $\Upsilon(\ul w;\tau,\sigma)$ kernels through the $S$ element, we simply have to identify $\tau\sim \tau+1$. For this reason, we will sometimes refer to the $STS$ element as $S'$ in order to emphasize that we are considering the $\mathbb{S}^3\times \mathbb{S}^1$ geometry. Also, in the case of $\mathbb{L}(r,1)\times \mathbb{S}^1$ there is another convenient parametrization of the fibration moduli and $g$ action. If we set
\be\label{q_tq_s}
q_\tau=\e^{2\pi\i\frac{\omega}{r\omega_1}}~,\quad q_\sigma=\e^{-2\pi\i\frac{\omega_3}{r\omega_1}}~,\quad \omega=\omega_1+\omega_2~,
\ee
and rescale $\bs X\to \bs X/r\omega_1$, $\bs \Xi\to \bs \Xi/r\omega_1$, then the $-g$ action simply exchanges $\omega_1\leftrightarrow \omega_2$ for any $r$. Geometrically, when $\omega_{1,2}$ are imaginary and positive they can be identified with the squashing parameters of $\mathbb{L}(r,1)$, while a real positive $\omega_3$ can be identified with the inverse of the $\mathbb{S}^1$ radius. In particular, for $r=1$ we have the identification (\ref{S3pq}) of parameters w.r.t. the previous subsection.

It is important to observe that the presence of the cubic polynomial $\mathcal{P}_3(\ul X,\ul \Xi)$ in (\ref{Z:PUU}) spoils the complete factorizability of the compact space integrand and large gauge invariance. However, when all the gauge and mixed-gauge anomalies vanish $\mathcal{P}_3(\ul X,\ul \Xi)$ is simply a constant, so that the whole integrand can be neatly factorized 
\be\label{Delta:UU}
\Delta_{\mathcal{M}_3\times \mathbb{S}^1}(\ul z,\ul \ell,\ul \zeta,\ul h)\propto \Upsilon(\ul w,\ul \xi;\tau,\sigma)^{(-g)}\;\Upsilon(\ul w,\ul \xi;\tau,\sigma)~.
\ee
This is an important point and we will come back to this problem in section \ref{sec:gaugeW}. Moreover, if the theory on $\mathbb{D}^2\times \mathbb{T}^2$ has isolated massive vacua only, which are in correspondence with the minima of the effective twisted superpotential $\mathcal{W}_{\rm eff}(\ul w)=-\lim_{\tau\to 0}\tau\ln\Upsilon(\ul w;\tau,\sigma)$, the whole compact space partition function is expected to factorize into {\it holomorphic/anti-holomorphic blocks } 
\be
{\rm Z}[\mathbb{D}^2\times\mathbb{T}^2]\propto \sum_{\gamma\in\rm vacua}\mathcal{B}_\gamma(\ul \xi;\tau,\sigma)^{(-g)}\;\mathcal{B}_\gamma(\ul \xi;\tau,\sigma)~,
\ee
where
\be
\mathcal{B}_\gamma(\ul \xi;\tau,\sigma)=\oint_{\gamma}\frac{\rd \ul w}{2\pi\i\ul w}\;\Upsilon(\ul w;\tau,\sigma)
\ee
is the block integral representation of the $\mathbb{D}^2\times \mathbb{T}^2$ partition function. This parallels the 3d setup discussed in \cite{Beem:2012mb}. Also, this result was checked in a few cases  by brute force computations \cite{Yoshida:2014qwa,Peelaers:2014ima,Chen:2014rca,Nieri:2015yia}, and the $W_{q,t;q'}(\Gamma)$ interpretation that we will give in the following sections indeed confirms that it should be generally valid. 

Following the same reasoning as in the previous subsection, the theory on the half-space can be enriched by the inclusion of 2d defects supported on $\mathbb{T}^2$, and the defect generating function reads
\be\label{D2T2genfun}
\mathcal{B}_\gamma(\{\tau_0^a,\ul\tau^a\};\tau,\sigma)=\e^{\mathcal{N}_0(\tau_0)}\oint_\gamma\frac{\rd \ul w}{2\pi\i\ul w}\;\Upsilon(\ul w;\tau,\sigma)\prod_a \exp\left(-\sum_{j=1}^{N_a}\sum_{n\neq 0}\frac{\tau^a_{n}w_{j,a}^n}{n(1-q_\sigma^n)}\right)~.
\ee
Since block integrals appear as elementary objects in the study of compact space partition functions, in the following section we will review their  $W_{q,t;q'}(\Gamma)$ algebra construction \cite{Nieri:2015dts}.

\section{Elliptic $W_{q,t;q'}$ algebras}\label{sec:ellW}
\begin{figure}[!ht]
\leavevmode
\begin{center}
\includegraphics[height=0.10\textheight]{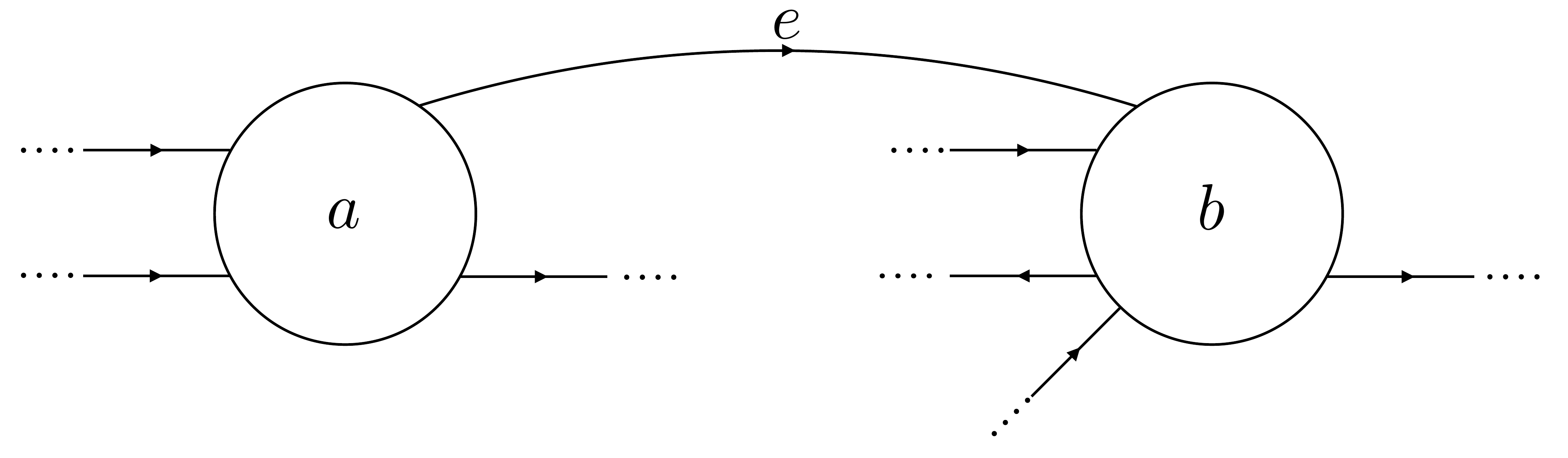}
\end{center}
\caption{Portion of a quiver $\Gamma$. We explicitly displayed two nodes $a,b\in\Gamma_0$, an arrow $e\in\Gamma_1$ from $a$ to $b$, and several arrows with source or target in $a$ or $b$.}
\label{Gamma}
\end{figure}

The elliptic deformation of the $W(A_1)$ (Virasoro) algebra was introduced in \cite{Nieri:2015dts} and generalized to arbitrary quiver diagrams in \cite{Kimura:2016dys}. Let us start by recalling some algebraic definitions from \cite{Kimura:2016dys}. A quiver $\Gamma$ is a collection of nodes $\Gamma_0$ and arrows $\Gamma_1$ (see figure \ref{Gamma} for an example).
Given two nodes $a,b\in\Gamma_0$ and arrows $\Gamma_1\ni e:a\to b$, we can associate to $\Gamma$ the deformed Cartan matrix $C_{ab}\in |\Gamma_0|\times |\Gamma_0|$ (not necessarily associated to a Lie algebra)
\be
C_{ab}=(1+p^{-1})\delta_{ab}-\sum_{e:b\to a}\mu_e^{-1}-p^{-1}\sum_{e:a\to b}\mu_e~,
\ee
and two commuting root-type Heisenberg algebras (we show non-trivial relations only)
\be\label{Heisell}
[\s^{(\pm)}_{a,n},\s^{(\pm)}_{b,m}]=\mp \frac{1-t^{\mp n}}{n(1-q^{\mp n})(1-q'^{\pm n})}C_{ab}^{[\pm n]}\delta_{n+m,0}~,\quad n,m\in\mathbb{Z}\backslash\{0\}~,\quad  [\s_{a,0},\tilde{\s}_{b,0}]=\beta C_{ab}^{[0]} ~,
\ee
where $\{q,t=q^\beta,p=qt^{-1},q',\mu_e\}\in\mathbb{C}$\footnote{To compare with \cite{Kimura:2016dys} we have to set $(q)_{\rm here}=(q_2)_{\rm there}$, $(t)_{\rm here}=(q_1^{-1})_{\rm there}$, $(q')_{\rm here}=(p)_{\rm there}$.}$^{,}$\footnote{The region of the parameter space may be restricted to ensure convergence of various expressions.}, while the ${}^{[n]}$ operation means replacing each parameter with its $n^{\rm th}$ power, for instance
\be
C_{ab}^{[n]}=(1+p^{-n})\delta_{ab}-\sum_{e:b\to a}\mu_e^{-n}-p^{-n}\sum_{e:a\to b}\mu_e^n~.
\ee
We next define the {\it screening current}
\be
\S^a(w)=\; :\e^{\sum_{n\neq 0}\left(\s_{a,n}^{(+)}w^{-n}+\s_{a,n}^{(-)}w^{n}\right)}:w^{\s_{a,0}}\e^{\tilde{\s}_{a,0}}~,
\ee
where the normal ordering symbol $:~:$ means moving all non-positive modes to the left. The elliptic $W_{q,t;q'}(\Gamma)$ algebra generated by $\{\T^a_{n},a\in\Gamma_0,n\in\mathbb{Z}\}$ can now be defined as the commutant of the screening charges $\J^a$ in the Heisenberg algebra, namely
\be\label{chargecons}
\J^a=\oint\frac{\rd w}{2\pi\i w}\;\S^a(w)~,\quad  [\T^a_{n},\J^b]=0 \!\quad \!\Leftrightarrow\! \quad\! [\T^a_{n},\S^b(w)]=\delta_{ab}\left(\O^b_{n}(q w)-\O^b_{n}(w)\right)~,
\ee
for some operator $\O^b_{n}(w)$ and integration contour. For the explicit form of $\O_n^b(w)$ see e.g. \cite{Kimura:2016dys}. The generating current $\T^a(w)=\sum_{n\in\mathbb{Z}}\T^a_n w^{-n}$ can be computed by means of the operator
\be
\Y^a(w)=\; :\e^{\sum_{n\neq 0}\left(\y_{a,n}^{(+)}w^{-n}+\y_{a,n}^{(-)}w^{n}\right)}:t^{\y_{a,0}-\tilde\rho_a}~,\quad \tilde\rho_a=\sum_{b\in\Gamma_0}\left(C^{-1}\right)^{[0]}_{ab}~,
\ee
where we have introduced the weight-type basis of the Heisenberg algebras defined by
\be
\begin{split}
&[\y_{a,n}^{(\pm)},\s_{b,m}^{(\pm)}]=\mp\frac{1-t^{\mp n}}{n(1-q'^{\pm n})}\delta_{n+m,0}\delta_{ab}~,\quad [\y_{a,0},\tilde{\s}_{b,0}]=[\s_{a,0},\tilde{\y}_{b,0}]=\delta_{ab}~,\\
&\y_{a,n}^{(\pm)}=(1-q^{\mp n})\left(C^{-1}\right)^{[\pm n]}_{ab}\s_{b,n}^{(\pm)}~,\quad \y_{a,0}=\beta^{-1}\left(C^{-1}\right)^{[0]}_{ab}\s_{b,0}~,\quad \tilde s_{a,0}=\tilde y_{b,0}\beta C_{ba}^{[0]}~.
\end{split}
\ee
The simplest example is provided by the single node quiver, in which case 
\be\label{ellVircurr}
\T(w)=\Y(w)+\Y(p^{-1} w)^{-1}~.
\ee
We refer to \cite{Kimura:2016dys} for the general construction. There is also a natural class of vertex operators given by
\be\label{Vertex}
\V^a(x)=\; :\e^{-\sum_{n\neq 0}\left(\frac{x^{-n}}{(1-t^{-n})(1-q^{n})}\y^{(+)}_{a,n}+\frac{x^{n}}{(1-t^{n})(1-q^{-n})}\y^{(-)}_{a,n}\right)}:~,\quad \H^a(x|u)=\; :\V^a(x)\V^a(x/qu)^{-1}:~,
\ee
of which we might eventually be interested in computing correlation functions.

\subsection{Matrix models}\label{sec:MatrixModels}
In this subsection we consider the matrix models which arise from $W_{q,t;q'}(\Gamma)$ algebras.\footnote{Another elliptic deformation of 2d CFT matrix models was proposed in \cite{Spiridonov:2011hf}.} These models are constructed by acting with an arbitrary number of screening charges on the charged Fock module $\mathcal{F}_{\ul\alpha}$ generated by the vacuum state $\ket{\ul\alpha}$ defined by
\be
\s^{(\pm)}_{a,n}\ket{\ul \alpha}=0~, \quad n>0~,\quad \ket{\ul \alpha}=\e^{\sum_{a}\alpha_a \tilde{\y}_{0,a}}\ket{0}~,\quad \s_{a,0}\ket{\ul \alpha}=\alpha_a\ket{\ul\alpha}~.
\ee
For later convenience we also recall that the dual Fock module is generated by $\bra{\ul\alpha}$, where
\be
\bra{\ul \alpha}\s^{(\pm)}_{a,-n}=0~, \quad n>0~,\quad \bra{\ul \alpha}=\e^{-\sum_{a}\alpha_a \tilde{\y}_{0,a}}\bra{0}~,\quad \bra{\ul \alpha}\s_{a,0}=\alpha_a\bra{\ul\alpha}~,\quad \braket{\ul\alpha}{\ul\alpha'}=\delta_{\ul\alpha,\ul\alpha'}~.
\ee
The resulting Fock state is (the product is over increasing indexes from left to right)
\be\label{Zstate}
\Z\ket{\ul\alpha}=\oint\frac{\rd\ul w}{2\pi\i\ul w}\;\prod_{a=1}^{|\Gamma_0|}\prod_{j=1}^{N_a}\S^a(w_{a,j})\ket{\ul\alpha}~.
\ee
A simple normal ordering computation yields
\begin{multline}
\S^a(w_{a,j})\S^b(w_{b,k})=\; :
\S^a(w_{a,j})\S^b(w_{b,k}):\times\\
\times \Delta^{(a)}_{\rm node}\left(\frac{w_{a,k}}{w_{a,j}}\right)^{\delta_{ab}} \!\!\! \Delta^{(a)}_{\rm self}\left(\frac{w_{a,k}}{w_{a,j}}\right)^{\delta_{ab}} \!\!\!  \Delta^{(ab)}_{\rm off}\left(\frac{w_{b,k}}{w_{a,j}}\right)^{1-\delta_{ab}} \!\!\! w_{b,k}^{-\beta C_{ba}^{[0]}}~,
\end{multline}
where
\begin{align}
\Delta^{(a)}_{\rm node}(w)&=\frac{\Gamma(t w;q,q')\Gamma(t w^{-1};q,q')}{\Gamma(w;q,q')\Gamma(w^{-1};q,q')}\frac{\Theta(t w^{-1};q)}{\Theta(w^{-1};q)}~,\\
\Delta^{(a)}_{\rm self}(w)&=\prod_{e:a\to a}\frac{\Gamma(\mu_e w;q,q')\Gamma(\mu_e w^{-1};q,q')}{\Gamma( t\mu_e w;q,q')\Gamma( t\mu_e w^{-1};q,q')}\frac{\Theta(\mu_e w^{-1};q)}{\Theta(t \mu_e w^{-1};q)}~,\\
\Delta^{(ab)}_{\rm off}(w)&=\prod_{e:a\to b}\frac{\Gamma(\mu_e w;q,q')}{\Gamma(t \mu_e w;q,q')}\prod_{e:b\to a}\frac{\Gamma(q t^{-1}\mu_e^{-1} w;q,q')}{\Gamma(q \mu_e^{-1} w;q,q')}~.
\end{align}
These functions determine the matrix model measure.
Let us introduce
\be
\Delta^{(a)}_{E}(\ul w_a)\!=\!\prod_{1\leq j\neq k\leq N_a}\frac{\Gamma(t w_{a,k}/w_{a,j};q,q')}{\Gamma(w_{a,k}/w_{a,j};q,q')}~,\quad \!\!\Delta^{(a)}_{\rm loop}(\ul w_a)\!=\!\prod_{e:a\to a}\prod_{1\leq j\neq k\leq N_a}\frac{\Gamma(\mu_e w_{a,k}/w_{a,j};q,q')}{\Gamma(t \mu_e w_{a,k}/w_{a,j};q,q')}~,
\ee
and the $q$-constants\footnote{This means that the function is invariant w.r.t. the shift $w_{a,j}\to q w_{a,j}$, i.e. it's an elliptic function.}
\be\label{qConst}
c^{(a)}_{\beta}(\ul w_a,\mu;q)\!=\!\prod_{1\leq j<k\leq N_a}\left(\frac{w_{a,j}}{w_{a,k}}\right)^\beta\frac{\Theta(t \mu w_{a,j}/w_{a,k};q)}{\Theta(\mu w_{a,j}/w_{a,k};q)}~,\quad \!\! C^{(a)}_\beta(\ul w_a;q)\!=\!\frac{c^{(a)}_\beta(\ul w_a,1;q)}{\prod_{e:a\to a} c^{(a)}_\beta(\ul w_a,\mu_e;q)}~.
\ee
Then the state (\ref{Zstate}) takes the form 
\be\label{matrixmodel1}
\Z\ket{\ul\alpha}=\oint\frac{\rd\ul w}{2\pi\i\ul w}\;\Delta_{\rm m.m.}(\ul w)\;
\prod_a \prod_j w_{a,j}^{-\beta((N_a-1)\frac{C_{aa}^{[0]}}{2}+ \sum_{a>b}C^{[0]}_{ab}N_b)}:\prod_{a}\prod_{j}\S_a(w_{a,j}):\ket{\ul\alpha}~,
\ee
where the matrix model measure is
\be\label{Deltamm}
\Delta_{\rm m.m.}(\ul w)=C^{(a)}_\beta(\ul w_a;q)\Delta^{(a)}_E(\ul w_a)\Delta^{(a)}_{\rm loop}(\ul w_a)\prod_{a<b}\prod_{j,k}\Delta^{(ab)}_{\rm off}\left(\frac{w_{b,k}}{w_{a,j}}\right)~.
\ee
Introducing the {\it time representation}
\be\label{timerep}
\boxed{~\begin{split}
\s_{a,-n}^{(\pm)}&\simeq \mp\frac{\tau^a_{\pm n}}{n(1-q'^{\pm n})}~,\quad s_{a,n}\simeq  \frac{1-t^{\mp n}}{1-q^{\mp n}}C_{ab}^{[\pm n]}\frac{\partial}{\partial \tau^b_{\pm n}}~,\quad n>0\\
\tilde{\y}_{a,0}&\simeq \tau^a_{0}~,\quad \s_{a,0}\simeq \frac{\partial}{\partial \tau^a_{0}}~,\quad \ket{0}\simeq 1
\end{split}~}~,
\ee
we arrive at the more conventional form of the matrix model
\be\label{matrixmodel2}
\boxed{~\Z\ket{\ul\alpha}\simeq {\rm Z}(\{\tau^a_0,\ul \tau^a\})=\e^{\mathcal{N}_0(\{\tau^a_0\})}\oint\frac{\rd \ul w}{2\pi\i\ul w}\;\Delta_{\rm m.m.}(\ul w)\;\e^{\sum_{a}\sum_{j=1}^{N_a} V^{(a)}(w_{a,j}|\ul\tau^a)}~}~,
\ee
where we identified the state
\be
\prod_a \prod_j w_{a,j}^{-\beta((N_a-1)\frac{C_{aa}^{[0]}}{2}+\sum_{a>b}C^{[0]}_{ab}N_b)}:\prod_{a}\prod_{j}\S_a(w_{a,j}):\ket{\ul\alpha}
\ee
with the exponential of the total matrix model potential given by
\be
V^{(a)}(w|\ul\tau^a)=\hat\alpha_a\ln w
-\sum_{n\neq 0} \frac{\tau^a_{n}w^n}{n(1-q'^{n})}~.
\ee
Here we set
\be
\hat\alpha_a=\alpha_a+\beta( \sum_{b}C_{ab}^{[0]}N_b-(N_a-1)\frac{C_{aa}^{[0]}}{2}-\sum_{a>b}C_{ab}^{[0]}N_b)~,
\ee
and defined the overall normalization
\be
\mathcal{N}_0(\{\tau^a_0\})=\sum_a \tau^a_{0}(\alpha_a+\beta\sum_b C_{ab}^{[0]}N_b)~.
\ee
It is worth noting that the matrix model can be enriched by the inclusion of additional vertex operators, for instance of the type (\ref{Vertex}). Their OPE with the screening current is  
\be
\V^a(x)\S^b(w)=\; : \V^a(x)\S^b(w):\;\Gamma(w/x;q,q')~,
\ee
and hence their contribution to the potential is equivalent to the shift
\be
V^{(a)}(w|\ul\tau^a)\to V^{(a)}(w|\ul\tau^a)+\sum_{n\neq 0}\frac{x^{-n} w^n}{n(1-q^n)(1-q'^n)}~,
\ee
corresponding to the following background for the time variables
\be
\tau^a_{n}\to \tau^a_{n}-\frac{x^{-n}}{1-q^n}~.
\ee

The $W_{q,t;q'}(\Gamma)$ construction of the matrix model (\ref{matrixmodel2}) allows us to write down a set of Ward identities that it has to satisfy. Here will give an explicit example based on the elliptic Virasoro ($W_{q,t;q'}(A_1)$) matrix model. In order to do so, we act with the elliptic Virasoro current (\ref{ellVircurr}) on the state (\ref{matrixmodel1}) with $|\Gamma_0|=1$, yielding 
\begin{multline}
\T(p^{1/2} z)\Z\ket{\ul\alpha}=\sum_{\sigma=\pm}\oint\frac{\rd\ul w}{2\pi\i\ul w}\;C_\beta(\ul w;q)\Delta_E(\ul w)
\prod_j w_{j}^{-\beta(N-1)}\prod_j f_\sigma(w_j/z)\times\\
\times:\Y(p^{\sigma/2}z)^\sigma\prod_{j}\S(w_{j}):\ket{\alpha}~,
\end{multline}
where 
\be
f_\sigma(w_j/z)=\frac{\Theta(p^{-\sigma/2}w_j/z;q')^\sigma}{\Theta(t^{-1}p^{-\sigma/2}w_j/z;q')^\sigma}~.
\ee
Using the time representation (\ref{timerep}), at the level of the matrix model (\ref{matrixmodel2}) we have
\be
T(p^{1/2}z){\rm Z}(\tau_0,\ul\tau)=\e^{\mathcal{N}_0(\tau_0)}\oint\frac{\rd \ul w}{2\pi\i\ul w}\;\Delta_{\rm m.m.}(\ul w)\left(\sum_{\sigma=\pm}\e^{Y_\sigma(z|\tau_0,\ul\tau)}\prod_{j=1}^N f_\sigma(w_j/z)\right)\e^{\sum_{j=1}^{N} V(w_{j}|\ul\tau)}~,
\ee
where the differential operator $T(z)$ is the time representation of elliptic Virasoro current, and $Y_\sigma(z|\tau_0,\ul\tau)$ is the potential arising from the non-positive modes of $\Y(p^{\sigma/2}z)^\sigma$. Similar Ward identities for general $W_{q,t;q'}(\Gamma)$ matrix models can in principle be derived using the explicit expression of the algebra generators given in \cite{Kimura:2016dys}.

\section{Elliptic $W_{q,t;q'}$ modular double}\label{sec:ellmodW}
In this section we show that when the deformation parameters of two independent elliptic $W_{q,t;q'}(\Gamma)$ algebras  are related by ${\rm SL}(2,\mathbb{Z})\subset {\rm SL}(3,\mathbb{Z})$ transformations, it is possible to combine them into a bigger algebra which we will call the elliptic modular double. The two independent commuting sectors will be also referred to as {\it chiral sectors}, in analogy with the usual 2d CFT terminology. In order to avoid possible confusion, we will use the index $i=1,2$ to exclusively denote the two chiral sectors and nothing else.

The construction essentially parallels \cite{Nedelin:2016gwu}, so we will simply recall the main ideas. Given two commuting elliptic algebras $W_{q,t;q'}(\Gamma)_{i=1,2}$ we have the associated screening currents $\S^a(w)_i$ defined by (\ref{chargecons})
\be\label{fundprop}
[\T^a_{n,i},\S^b(w)_{i'}]=\delta_{ii'}\delta_{ab}\left(\O_{n}(q_i w)_i-\O_{n}(w)_i\right)~.
\ee
Let us now parametrize the positions through
\be
(w)_i=\e^{\frac{2\pi\i}{\omega_i}X}~,
\ee
and let us tentatively define the operator 
\be\label{Smod}
\mathcal{S}^a(X)=\S^a(w)_1\otimes\; \S^a(w)_2=\otimes_{i=1,2}[\S^a(w)_i]_-[\S^a(w)_i]_+[\S^a(w)_i]_0~,
\ee
where $[~]_{\pm,0}$ denotes the positive/negative oscillator parts and the zero modes respectively. In general, this operator is not a screening current for any $W_{q,t;q'}(\Gamma)_i$ since
\be
[\T^a_{n,1},\mathcal{S}^b(X)]=\delta_{ab}\left(\O^b_{n}(q_1w)_1-\O^b_{n}(w)_1\right)\otimes \S^b(w)_2\neq \textrm{total difference}~,
\ee
and similarly for the other sector. However, in the special case when  
\be
q_i=\e^{2\pi\i\frac{\omega}{\omega_i}}~,\quad q_i'=\e^{-2\pi\i\frac{\omega_3}{\omega_i}}~,\quad t_i=\e^{2\pi\i\beta\frac{\omega}{\omega_i}}~,\quad \mu_{e,i}=\e^{2\pi\i\frac{M_e}{\omega_i}}~,\quad \omega=\omega_1+\omega_2~,
\ee
we see that, if we assume $\omega_i$-periodicity in the $i^{\rm th}$ chiral sector, we could write 
\be
[\T^{a}_{n,1},\mathcal{S}^b(X)]=\delta_{ab}\left(\mathcal{O}^b_{n}(\omega_{2}+X)-\mathcal{O}^b_{n}(X)\right)=\textrm{total difference}~,
\ee
with $ \mathcal{O}^b_{n}(X)=\O^b_{n}(w)_1\otimes \S^b(w)_2$, and similarly for the other sector. The only obstruction to this reasoning is given by the non-trivial monodromy of the zero modes $[\S^b(w)_i]_0=(w)_i^{\s_{0,b,i}}$ under the shift $X\to X+\omega_i$. Here we propose the following solution, also mentioned in \cite{Nedelin:2016gwu} without giving the details. We replace the zero mode part in the definition (\ref{Smod}) with the following operator\footnote{We can define this operator through the series expansion $\frac{\Theta(w x;q)}{\Theta(w;q)\Theta(x;q)}=\frac{1}{(q;q)^2_\infty}\sum_{n\in\mathbb{Z}}\frac{x^n}{1-w q^n}$.}
\be
[\S^a(w)_i]_0 \longrightarrow \frac{\Theta((w)_i \; q_i^{-\s_{0,a,i}};q_i)}{ \Theta((w)_i;q_i)\Theta(q_i^{-\s_{0,a,i}};q_i)}~,
\ee
which does not change the fundamental property (\ref{fundprop}) because 
\be
(w)_i^{-\s_{0,a,i}} \frac{\Theta((w)_i \; q_i^{-\s_{0,a,i}};q_i)}{\Theta((w)_i;q_i)\Theta(q_i^{-\s_{0,a,i}};q_i)}=q_i\textrm{-constant}~,
\ee
namely this combination is invariant w.r.t. $X\to X+\omega_i$. However, now there are no monodromies and $\mathcal{S}^a(X)$ is a simultaneous screening current for both $W_{q,t;q'}(\Gamma)_{i=1,2}$. We can call the resulting algebra $W_{q,t;q'}^{g}(\Gamma)$ for short, with $g=S'=STS$ in this case. The name arises from the fact that in the given parametrization the two chiral sectors are related by $\omega_1\leftrightarrow\omega_2$, and if we reparametrize 
\be
\tau=\frac{\omega}{\omega_1}~,\quad \sigma=-\frac{\omega_3}{\omega_1}~,
\ee
and rescale $X/\omega_1\to X$, $M_e/\omega_1\to M_e$, then the exchange $\omega_1\leftrightarrow\omega_2$ corresponds to the $S'\in{\rm SL}(2,\mathbb{Z})\subset {\rm SL}(3,\mathbb{Z})$ action we already encountered in section \ref{sec:gauge} 
\be
-S':\tau \mapsto \frac{1}{\tau-1}~,\quad -S':\sigma \mapsto \frac{\sigma}{\tau-1}~,\quad -S': M_e \mapsto \frac{M_e}{\tau-1}~,\quad -S': X \mapsto \frac{X}{\tau-1}~.
\ee

In fact, it turns out that this construction can be extended to other ${\rm SL}(2,\mathbb{Z})\subset {\rm SL}(3,\mathbb{Z})$ elements. The parameters in the $1^{\rm st}$ and $2^{\rm nd}$ chiral sectors are 
\be\label{cp1}
~~~~~~~~~\begin{array}{c|c|c||c|c|c}
~i=1~&~W_{q,t;q'}^{ST^rS}(\Gamma)~&~W_{q,t;q'}^{id}(\Gamma)&~i=2~&~W_{q,t;q'}^{ST^rS}(\Gamma)~&~W_{q,t;q'}^{id}(\Gamma)\\
\hline
q_1&\multicolumn{2}{c||}{\e^{2\pi\i\tau}}&q_2&\multicolumn{2}{c}{\e^{-2\pi\i g\cdot\tau}}\\
\hline
q'_1&\multicolumn{2}{c||}{\e^{2\pi\i\sigma}}&q'_2&\multicolumn{2}{c}{\e^{-2\pi\i g\cdot\sigma}}\\
\hline
t_1&\multicolumn{2}{c||}{\e^{2\pi\i\beta_1\tau}}&t_2&\multicolumn{2}{c}{\e^{-2\pi\i\beta_2 g\cdot\tau}}\\
\hline
\mu_{e,1}&\multicolumn{2}{c||}{\e^{2\pi\i M_e}}&\mu_{e,2}&\multicolumn{2}{c}{\e^{-2\pi\i  g\cdot M_e}}\\
\hline
(w)_1~&~\e^{2\pi\i X}\e^{\frac{2\pi\i}{r}\ell}~&~\e^{2\pi\i X}\e^{-\i\pi\tau\ell}&(w)_2~&~\e^{-2\pi\i g\cdot X}\e^{-\frac{2\pi\i}{r}g\cdot \ell}~&~\e^{-2\pi\i g\cdot X}\e^{-\i\pi g\cdot \tau \ell}\\
\end{array}~~~~~,
\ee
and the gluing is according to the following $g\in {\rm SL}(2,\mathbb{Z})\subset {\rm SL}(3,\mathbb{Z})$ action ($r\in\mathbb{Z}_{>0}$)
\be\label{cp2}
~~~~~~\begin{array}{c|c|c}
~&~W_{q,t;q'}^{ST^rS}(\Gamma)~&~W_{q,t;q'}^{id}(\Gamma)\\
\hline
-g\cdot \tau~&~\frac{\tau}{r\tau-1}~&~-\tau\\
\hline
-g\cdot \sigma~&~\frac{\sigma}{r\tau-1}~&~-\sigma\\
\hline
\beta_2~&~\beta_1~&~-\beta_1\\
\hline
-g\cdot M_e~&~\frac{M_e}{r\tau-1}~&~M_e\\
\hline
-g\cdot (X,\ell)~&~(\frac{ X}{r\tau-1},r-\ell)~&~ (X,\ell)
\end{array}\quad\quad~.
\ee
Summarizing, the $W_{q,t;q'}^g(\Gamma)$ screening currents are given by
\be\label{Smodd}
\boxed{~\mathcal{S}^a(X)=\sum_{\ell\in\mathbb{F}}\otimes_{i=1,2}[\S^a(w)_i]_-[\S^a(w)_i]_+ \frac{\Theta((w)_i \; q_i^{-\s_{0,a,i}};q_i)}{\Theta((w)_i;q_i)\Theta(q_i^{-\s_{0,a,i}};q_i)}~}~,
\ee
where $\mathbb{F}=(\emptyset,\mathbb{Z}_r,\mathbb{Z})$ for $g=(S',ST^r S,id)$ respectively. Also, the modular double version of the vertex operators (\ref{Vertex}) for $g=S'$ is
\be
\mathcal{V}^a(\chi)=\; \V^a(x)_1\otimes \V^a(x)_2~,\quad \mathcal{H}^a(\chi|U)=\; :\mathcal{V}^a(\chi)\mathcal{V}^a(\chi-\omega-U)^{-1}:~,
\ee
where we have defined
\be
(x)_i=\e^{2\pi\i\frac{\chi}{\omega_i}}~,\quad (u)_i=\e^{2\pi\i\frac{U}{\omega_i}}~.
\ee
A similar definition can be given for the other elements.

\subsection{Matrix models}
Given the screening current (\ref{Smodd}) it is possible to build the associated matrix model. Since everything parallels what we have already discussed in subsection \ref{sec:MatrixModels} for a single chiral copy, we will be brief.  The matrix model of the modular double is constructed by acting with an arbitrary number of screening charges on the Fock module $\mathcal{F}_{\ul\alpha}$ generated by the vacuum state $\ket{\ul\alpha}$ defined by
\be
\s^{(\pm)}_{a,n,i} \ket{\ul \alpha}=0~, \quad n>0~,\quad \ket{\ul \alpha}=\e^{\sum_{a}\alpha_a \tilde{\y}_{0,a,i}}\ket{0}~,\quad \s_{a,0,i}\ket{\ul \alpha}=\alpha_a\ket{\ul\alpha}~,\quad i=1,2.
\ee
For later convenience we also recall that the dual Fock module is generated by $\bra{\ul\alpha}$, where
\be
\bra{\ul \alpha}\s^{(\pm)}_{a,-n,i} =0~, \quad n>0~,\quad \bra{\ul \alpha}=\bra{0}\e^{-\sum_{a}\alpha_a \tilde{\y}_{0,a,i}}~,\quad \bra{\ul \alpha}\s_{a,0,i}=\alpha_a\bra{\ul\alpha}~,\quad\!\!\! \braket{\ul\alpha}{\ul\alpha'}=\delta_{\ul\alpha,\ul\alpha'}~,\!\!\!\quad i=1,2~.
\ee
The resulting Fock state is (the product is over increasing indexes from left to right)
\be
\mathcal{Z}_g\ket{\ul\alpha}=\int\!\!\!\rd \ul X\;\prod_{a=1}^{|\Gamma_0|}\prod_{j=1}^{N_a}\mathcal{S}^a(X_{a,j})\ket{\ul\alpha}~.
\ee
Using (\ref{matrixmodel2}) we can conclude that this state yields the matrix model
\be\label{modmatrixmodel}
\boxed{~\mathcal{Z}_g(\{\tau^a_{0,i},\ul \tau^a_i\})=\e^{\sum_{i=1,2}\mathcal{N}_0(\{\tau^a_{0,i}\})}\!\!\!\!\!\sum_{\{\ul\ell\in\mathbb{F}^{N_a}\}}\int\!\!\!\rd \ul X\;\Delta_g(\ul X,\ul\ell)\;\e^{\sum_{i=1,2}\sum_{a}\sum_{j=1}^{N_a} \mathcal{V}_i^{(a)}(X_{a,j},\ell_{a,j}|\ul\tau^a_i)}~}~,
\ee
where the measure is given by the product of two copies of the measure (\ref{Deltamm}) related by the $g\in {\rm SL}(2,\mathbb{Z})\subset {\rm SL}(3,\mathbb{Z})$ action given in tables (\ref{cp1}), (\ref{cp2})
\be\label{Delta:DmDm}
\boxed{~
\Delta_g(\ul X,\ul \ell)=\Delta_{\rm m.m.}(\ul w)^{(-g)}\Delta_{\rm m.m.}(\ul w)~}~,
\ee
while the potential is given by
\be
\mathcal{V}_i^{(a)}(X ,\ell |\ul\tau^a_i)=\hat\alpha_a\ln (w)_i
-\sum_{n\neq 0} \frac{\tau^a_{n,i} (w^n)_i}{n(1-q_i'^{n})}~.
\ee
In this computation, we used the modular properties (\ref{ThetaS}), (\ref{ThetaSr}), (\ref{ThetaA}) of $\Theta$ functions to deal with the terms coming from the zero modes. Following the discussion of subsection \ref{sec:MatrixModels}, we can also conclude that the matrix model (\ref{modmatrixmodel}) will satisfy two independent sets of Ward identities related by $g\in {\rm SL}(2,\mathbb{Z})\subset {\rm SL}(3,\mathbb{Z})$.

In the next section we will discuss the relationship between elliptic $W_{q,t;q'}(\Gamma)$ matrix models and gauge theory partition functions from section \ref{sec:gauge}.

\section{Gauge/$W_{q,t;q'}$ correspondence}\label{sec:gaugeW}
The similarities between tables (\ref{ggluing}) and (\ref{cp1}), (\ref{cp2}) upon suitable identification of parameters is no coincidence. In this section we show that the (chiral) elliptic $W_{q,t;q'}(\Gamma)$ matrix model (\ref{matrixmodel2}) reproduces the holomorphic blocks and generating function of an associated gauge theory on the half-space $\mathbb{D}^2\times\mathbb{T}^2$, while the elliptic $W_{q,t;q'}^g(\Gamma)$ matrix model (\ref{modmatrixmodel}) reproduces the partition function and generating function of the gauge theory on the compact space $\mathcal{M}_3\times \mathbb{S}^1$ obtained from $\mathbb{D}^2\times\mathbb{T}^2$ through the $g\in {\rm SL}(2,\mathbb{Z})\subset {\rm SL}(3,\mathbb{Z})$ gluing.

\subsection{Half-space/chiral matrix model}
As discussed in subsection \ref{sec:d2xt2}, 4d holomorphic blocks can be interpreted as gauge theory partition functions on $\mathbb{D}^2\times\mathbb{T}^2$. For a gauge group $G$, they can be given in terms of contour integrals of a meromorphic kernel along a middle dimensional cycle $\gamma\subset (\mathbb{C}^\times)^{|G|}$ 
\be
\mathcal{B}_\gamma(\ul \xi;\tau,\sigma)=\oint_{\gamma}\frac{\rd \ul w}{2\pi\i\ul w}\;\Upsilon(\ul w;\tau,\sigma)~,
\ee
where $\ul \xi$ are fugacities for global symmetries and $\tau$, $\sigma$ are the disk equivariant parameter and torus modulus respectively. In order to establish a connection with elliptic $W_{q,t;q'}(\Gamma)$ algebras we have to restrict to {\it unitary or special unitary gauge groups}. For simplicity, let us start by discussing the first possibility, in which case we have $G=\bigtimes_{a}{\rm U}(N_a)$. The integral kernel depends on the specific theory and can be assembled using the elementary building blocks given in (\ref{Z1loop:d2xt2}). We will consider a theory with $\mathcal{N}=2$ multiplet content (superpotential couplings may be more general). Upon suitable identification of parameters, it is not difficult to recognize in the elliptic matrix model measure (\ref{Deltamm}) the total integral kernel of a gauge theory whose quiver diagram is dictated by the quiver $\Gamma$ of the algebra
\be
\boxed{~\Upsilon(\ul w;\tau,\sigma)=\Delta_{\rm m.m.}(\ul w)~}~.
\ee
\begin{figure}[!ht]
\leavevmode
\begin{center}
\includegraphics[height=0.14\textheight]{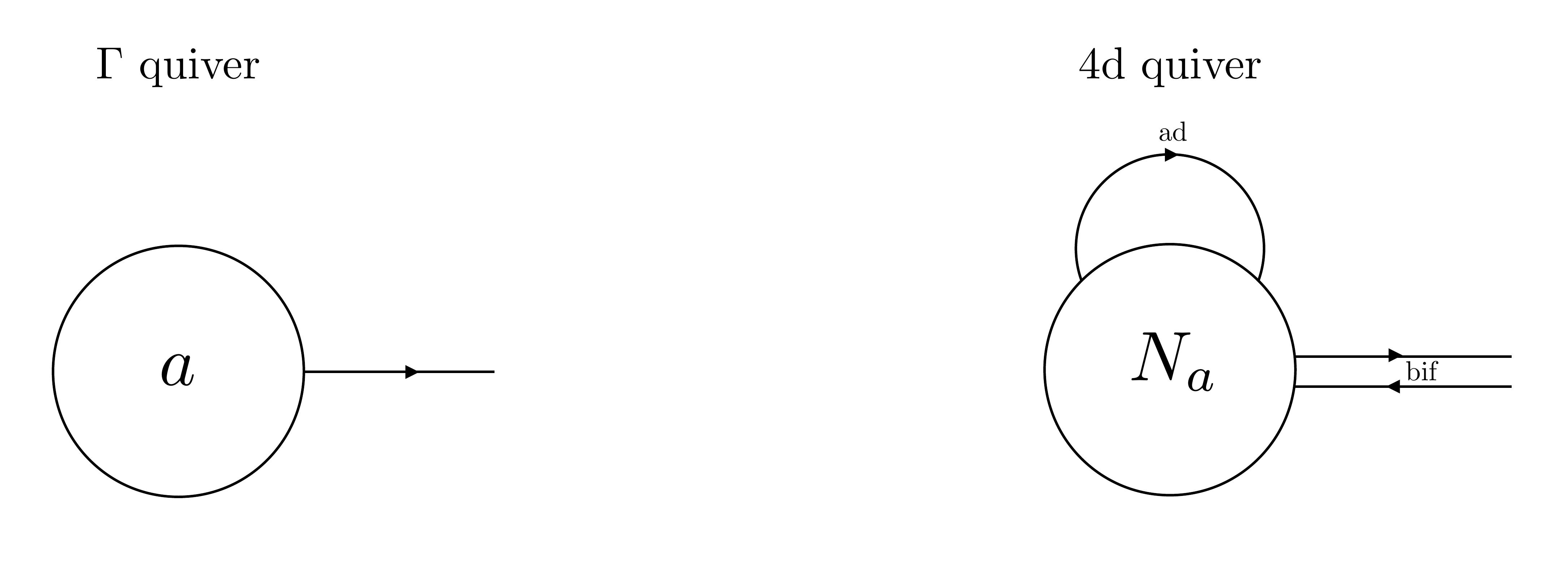}
\end{center}
\caption{Portion of a $\Gamma$ quiver (left) and the associated gauge theory quiver (right).}
\label{4dquiver}
\end{figure}
In particular, the number $N_a$ of screening currents of each kind is identified with the rank of the gauge group associated to the node $a\in\Gamma_0$ and the $\ul w$ coordinates of the screening currents with the gauge variables. Each node $a\in \Gamma_0$ is associated to a ${\rm U}(N_a)$ vector multiplet together with an adjoint chiral (in fact, one $\mathcal{N}=2$ vector) whose contribution to the kernel is 
\be
\Upsilon^{(a)}_{\rm vec}(\ul w_a)=\prod_{1\leq j\neq k\leq N_a}\frac{\Gamma(t_a w_{a,j}/w_{a,k};q_\tau,q_\sigma)}{\Gamma(w_{a,j}/w_{a,k};q_\tau,q_\sigma)}~.
\ee
Each oriented arrow $\Gamma_1 \ni e:a\to b$ (or $e:b\to a$) between different nodes ($a<b$) is associated to a pair of bifundamental chirals (in fact, one $\mathcal{N}=2$ bi-fundamental hyper) contributing with
\be
\begin{split}\Upsilon^{(e:a\to b)}_{\rm bif}(\ul w_a,\ul w_b)&=\prod_{j=1}^{N_a}\prod_{k=1}^{N_b}\frac{\Gamma(\xi_{e} w_{b,k}/w_{a,j};q_\tau,q_\sigma)}{\Gamma(q_\tau\bar{\xi}_{e} w_{b,k}/w_{a,j};q_\tau,q_\sigma)}~,\\
\Upsilon^{(e:b\to a)}_{\rm bif}(\ul w_a,\ul w_b)&=\prod_{j=1}^{N_a}\!\prod_{k=1}^{N_b}\frac{\Gamma(\bar\xi_{e}^{-1} w_{b,k}/w_{a,j};q_\tau,q_\sigma)}{\Gamma(q_\tau \xi_{e}^{-1} w_{b,k}/w_{a,j};q_\tau,q_\sigma)}~,
\end{split}
\ee
while each loop arrow $\Gamma_1\ni e:a\to a$ is associated to a pair of adjoint chirals (in fact, one $\mathcal{N}=2$ adjoint hyper) whose contribution is 
\be
\Upsilon^{(a)}_{\rm ad}(\ul w_a)=\prod_{1\leq j\neq k\leq N_a}\frac{\Gamma( \xi_{a} w_{a,j}/w_{a,k};q_\tau,q_\sigma)}{\Gamma(q_\tau \bar \xi_{a} w_{a,j}/w_{a,k};q_\tau,q_\sigma)}~.
\ee
These gauge theory data determine the algebra and vice-versa. The fugacities and deformation parameters of the algebra are easily identified
\be\label{gauge:algebra1}
\begin{array}{c|c|c|c|c|c|c}
~\textrm{Gauge theory}~&~q_\tau~&~q_\sigma~&~t_{a}~&~w_{a,j}~&\xi_{e}~&~\xi_e/\bar\xi_{e}\\
\hline
~\textrm{$W_{q,t;q'}(\Gamma)$}~&~q~&~q'~&~t=q^\beta~&~w_{a,j}~&~\mu_{e}~&~p=q t^{-1}
\end{array}\quad~.
\ee

Moreover, from the gauge theory perspective there can be FI parameters for each ${\rm U}(1)$ factor contributing with
\be
\Upsilon^a_{\rm FI}(\ul w_a)=\prod_{j=1}^{N_a}w_{a,j}^{\kappa_a}~,
\ee
and pairs of fundamental/anti-fundamental chirals (in fact, one $\mathcal{N}=2$ hyper) contributing with
\be
\Upsilon^a_{\rm f}(\ul w_a)=\prod_{j=1}^{N_a}\prod_{f\geq 1}\frac{\Gamma(w_{a,j} /\bar \xi_{a,f} ;q_\tau,q_\sigma)}{\Gamma(q_\tau w_{a,j} /  \xi_{a,f};q_\tau,q_\sigma)}~.
\ee
These additional data will appear on the algebra side when evaluating particular states or correlators of vertex operators $\H^a(x|u)$ of the type (\ref{Vertex}), and the correspondence is
\be\label{gauge:algebra2}
\begin{array}{c|c|c|c}
~\textrm{Gauge theory}~&~\bar \xi_{a,f}~&~\bar \xi_{a,f}/\xi_{a,f}~&~\kappa_a~\\
\hline
~\textrm{$W_{q,t;q'}(\Gamma)$}~&~x_{a,f}~&~u_{a,f}~&~\hat\alpha_a~
\end{array}\quad~.
\ee

The only factors which remain to be discussed are the $q$-constants (\ref{qConst}). Since the integrals are computed by residues and the integration contours are chosen to enclose poles for which $w_{a,j}/w_{a,k}\propto q^{\mathbb{Z}}$, it turns out that $q$-constants can actually be pulled out of the integral. Then we can identify 
\be\label{holblock:corr}
\boxed{~
\mathcal{B}_\gamma(\ul \xi,\bar{\ul\xi};\tau,\sigma)\propto \bra{\ul \alpha_\infty}\prod_{a}\prod_{f}\H^a(x_{a,f}|u_{a,f})\;\Z\ket{\ul\alpha}~}~,
\ee
where $(\alpha_\infty)_a=\alpha_a+\beta C_{ab}^{[0]}N_b$. The correspondence can actually be extended to the generating function (\ref{D2T2genfun}), which is identified with the state
\be\label{holblock:gen}
\boxed{~\mathcal{B}_\gamma(\{\tau^a_0,\ul\tau^a\};\tau,\sigma)\simeq \Z\ket{\ul\alpha}\simeq {\rm Z}(\{\tau^a_0,\ul\tau^a\})~}~.
\ee
The holomorphic block/correlator (\ref{holblock:corr}) then corresponds to the specialization/projection 
\be
\tau^a_{n}=-\sum_f \frac{1-q^{n}u_{a,f}^{n}}{1-q^n}x^{-n}_{a,f}~.
\ee

The $W_{q,t;q'}(\Gamma)$ interpretation of the partition (generating) function of special unitary gauge theories deserves additional comments. For simplicity, let us focus on the single node theory. From the gauge theory viewpoint, it is useful to integrate over the ${\rm SU}(N)$ algebra by considering the ${\rm U}(N)$ measure with a $\delta$ function constraint enforcing the traceless condition on the generators, for instance 
\be
\mathcal{B}_\gamma[{\rm U}(N)]=\oint_{\gamma}\frac{\rd^N \ul w}{2\pi\i\ul w}\;\Upsilon(\ul w;\tau,\sigma)~\longrightarrow ~\mathcal{B}_\gamma[{\rm SU}(N)]=\oint_{\gamma}\frac{\rd^{N} \ul w}{2\pi\i\ul w}\;\hat\delta(\prod_j w_j)\Upsilon(\ul w;\tau,\sigma)~,
\ee
where $\hat\delta(\prod_j w_j)$ removes one integration and imposes $\prod_j w_j=1$. Because of the relations (\ref{holblock:corr}), (\ref{holblock:gen}), on the $W_{q,t;q'}(\Gamma)$ algebra side this means that we are effectively modifying the $\Z$ operator according to  
\be
\Z=\oint \frac{\rd^N \ul w}{2\pi\i\ul w}\;\prod_j \S(w_j)~\longrightarrow ~ \Z_o=\oint \frac{\rd^N \ul w}{2\pi\i\ul w}\;\hat\delta(\prod_j w_j)\prod_j \S(w_j)~,
\ee
constraining the positions of the screening currents to be not all independent but rather to lie on the locus $\prod_j w_j=1$. However, from the $W_{q,t;q'}(\Gamma)$ perspective this operation may be not completely straightforward because it raises analytical questions which cannot be easily addressed through algebraic manipulations only. Here we propose another approach to ${\rm SU}(N)$ matrix models. We keep the operator $\Z$ untouched, but we evaluate its matrix elements (with additional insertions, if any)
\be
\bra{\rm out}\;\cdots\; \Z\ket{\rm in}
\ee
on different states. Usually, the $\ket{\rm in}$ and $\bra{\rm out}$ Fock states are chosen to be eigenstates of the momentum operator $\s_0$, namely
\be
\s_0\ket{\alpha}=\alpha\ket{\alpha}~,\quad \bra{\alpha}\s_0=\alpha\bra{\alpha}~,\quad \ket{\alpha}=\e^{\alpha\tilde\y_0}\ket{0}~,\quad \bra{\alpha}=\bra{0}\e^{-\alpha\tilde\y_0}~.
\ee  
Instead, we can evaluate the matrix elements of $\Z$ between eigenstates  of the coordinate operator $\tilde\y_0$, which in $\ket{\alpha}$ basis read
\be
\ket{\tau_0}=\sum_{\alpha}\ket{\alpha}\;\e^{-\alpha\tau_0}~,\quad  \braket{\alpha}{\tau_0}=\e^{-\alpha\tau_0}~,\quad \bra{\tau_0}=\sum_{\alpha}\bra{\alpha}\;\e^{\alpha\tau_0}~,\quad \braket{\tau_0}{\alpha}=\e^{\alpha\tau_0}~.
\ee
Then we are lead to consider the dressed operator
\be
\Z_o=\sum_{\alpha,\alpha'}\e^{(\alpha'-\alpha)\tau_0}\e^{-\alpha'\tilde \y_0}\;\Z\;\e^{\alpha\tilde \y_0}~,
\ee
and its expectation value between ground states. Because of charge conservation, we will get non-trivial results only when $\alpha'=\alpha+2\beta N$, implying
\be
\bra{0}\;\cdots\;\Z_o\ket{0}=\e^{2\beta N\tau_0}\bra{0}\;\cdots\;\sum_\alpha \e^{-(\alpha+2\beta N)\tilde\y_0}\;\Z\;\e^{\alpha\tilde\y_0}\ket{0}~.
\ee
As we know from the computations of subsection \ref{sec:MatrixModels}, the state under summation has a momentum dependence of the form 
\be
\oint\frac{\rd^N\ul w}{2\pi\i\ul w}\;\cdots \;(\prod_j w_j)^\alpha\; \ket{0}~,
\ee 
and taking the summation over $\alpha\in\mathbb{Z}$ we get the following representation of the $\delta$ function\footnote{Over periodic functions of $X\sim X+1$, $w=\e^{2\pi\i X}$, otherwise we integrate over a continuos momentum.}
\be
\hat\delta(\prod_j w_j)=\sum_{\alpha\in\mathbb{Z}}(\prod_j w_j)^\alpha~.
\ee
It seems that the algebraic properties of special unitary matrix models have not been studied as much extensively as their unitary cousins in the literature. The approach that we have outlined here is also applicable to the more conventional rational and trigonometric models and it deserves further investigations, especially in view of their gauge theory applications.

\subsection{Compact space/modular double matrix model}
Due to the equivalence of table (\ref{ggluing}) with (\ref{cp1}), (\ref{cp2}) through the dictionaries (\ref{gauge:algebra1}), (\ref{gauge:algebra2}) worked out in the previous subsection, it turns out that the gauge theory kernel (\ref{Delta:UU}) associated to the compact space $\mathcal{M}_3\times \mathbb{S}^1\simeq [\mathbb{D}^2\times\mathbb{T}^2]\cup_g [\mathbb{D}^2\times\mathbb{T}^2]$ can be identified with the measure (\ref{Delta:DmDm}) of the elliptic $W_{q,t;q'}^g(\Gamma)$ matrix model 
\be
\boxed{~
\Delta_{~\mathcal{M}_3\times S^1}(\ul z,\ul \ell)\propto \Upsilon(\ul w;\tau,\sigma)^{(-g)}\Upsilon(\ul w;\tau,\sigma)=\Delta_{\rm m.m.}(\ul w)^{(-g)}\Delta_{\rm m.m.}(\ul w)=\Delta_{g}(\ul X,\ul \ell)~}~.
\ee
The correspondence can actually be extended to the whole generating function/$W_{q,t;q'}^g(\Gamma)$ matrix model, but in order to establish a precise correspondence with the results reviewed in section \ref{sec:gauge} some work is necessary. Indeed, we have not explained how the $(\ldots)^{(-g)}(\ldots)$ factorization of the compact space measure can be achieved. In this section we review how this works for the $\mathbb{S}^3\times \mathbb{S}^1$ geometry, namely the element $S'=STS\in{\rm SL}(2,\mathbb{Z})\subset {\rm SL}(3,\mathbb{Z})$, and refer to \cite{Nieri:2015yia} for the other backgrounds.

Let us start by discussing the bare partition function (\ref{DefGenTau}), namely the generating function at $\ul\tau^a=0$. In notation of section \ref{sec:gauge}, the $\mathbb{S}^3\times \mathbb{S}^1$ partition function of the theory associated to the quiver $\Gamma$ with gauge group $G=\bigtimes_{a}{\rm U}(N_a)$ reads (up to normalization)
\be
{\rm Z}[\mathbb{S}^3\times \mathbb{S}^1]=\oint_{\mathbb{T}^{|G|}}\frac{\rd \ul z}{2\pi\i\ul z}\;\Delta_{\mathbb{S}^3\times \mathbb{S}^1}(\ul z)~,
\ee
where we defined
\begin{align}
\Delta_{\mathbb{S}^3\times \mathbb{S}^1}(\ul z)&=\Delta_{2\,\rm vec}(\ul z)\Delta_{\rm ad}(\ul z)\Delta_{\rm bif}(\ul z)\Delta_{\rm FI}(\ul z)\Delta_{\rm f}(\ul z)~,\nn\\
\Delta_{2\,\rm vec}(\ul z)&=\prod_{a}\!\prod_{1\leq j\neq k\leq N_a}\!\!\!\frac{\Gamma(\hat\t z_{a,j}/z_{a,k};\p,\q)}{\Gamma(z_{a,j}/z_{a,k};\p,\q)}~,\quad \!\!\!\Delta_{\rm ad}(\ul z)=\prod_{a}\!\prod_{e:a\to a}\prod_{1\leq j\neq k\leq N_a}\!\!\!\frac{\Gamma(\mu_e z_{a,j}/z_{a,k};\p,\q)}{\Gamma(\p\q \bar \mu_e z_{a,j}/z_{a,k};\p,\q)}~,\nn\\
\Delta_{\rm bif}(\ul z)&=\prod_{a<b}\prod_{e:a\to b}\prod_{j=1}^{N_a}\prod_{k=1}^{N_b}\frac{\Gamma(\mu_e z_{b,k}/z_{a,j};\p,\q)}{\Gamma(\p\q \bar \mu_e z_{b,k}/z_{a,j};\p,\q)}\prod_{e:b\to a}\prod_{j=1}^{N_a}\prod_{k=1}^{N_b}\frac{\Gamma( \bar\mu_e^{-1} z_{b,k}/z_{a,j};\p,\q)}{\Gamma(\p\q\mu_e^{-1} z_{b,k}/z_{a,j};\p,\q)}~,\nn\\
\Delta_{\rm FI}(\ul z)&=\prod_{a}\prod_{j=1}^{N_a} z_{a,j}^{\kappa_a}~,\quad \Delta_{\rm f}(\ul z)=\prod_{a}\prod_{f}\frac{\Gamma( z_{a,j}/\bar \mu_f;\p,\q)}{\Gamma(\p\q  z_{a,j}/ \mu_f;\p,\q)}~.
\end{align}
The total integrand has the same structure as the half-space kernel $\Upsilon(\ul w;\tau,\sigma)$ considered in the previous subsection, but it does not look like $\Upsilon(\ul w;\tau,\sigma)^{(-S')}\Upsilon(\ul w;\tau,\sigma)$ yet. However, let us recall from section \ref{sec:gauge} the following parametrization adapted to the $\mathbb{S}^3\times \mathbb{S}^1$ geometry
\be
\q=\e^{2\pi\i\frac{\omega_1}{\omega_3}}~,\quad \p=\e^{2\pi\i\frac{\omega_2}{\omega_3}}~,
\ee
and
\be
z_{a,j}=\e^{\frac{2\pi\i }{\omega_3}X_{a,j}}~\!,\!\quad \!\hat\t=\e^{\frac{2\pi\i }{\omega_3}\hat T}\!~,\!\quad\! \mu_e=\e^{\frac{2\pi\i }{\omega_3}M_e}\!~,\!\quad \!\bar \mu_e=\e^{\frac{2\pi\i }{\omega_3}\bar M_e}\!~,\!\quad\! \mu_f=\e^{\frac{2\pi\i }{\omega_3}M_f}\!~,\!\quad \!\bar\mu_f=\e^{\frac{2\pi\i }{\omega_3}\bar M_f}\!~.
\ee
The advantage of this parametrization is that the $S'$-gluing corresponds to the exchange $\omega_1\leftrightarrow \omega_2$ and naturally extends to the $ST^rS$-gluing for $\mathbb{L}(r,1)\times \mathbb{S}^1$. Using the modular property (\ref{gammamod}) 
it is possible to rewrite each elliptic $\Gamma$ function appearing in the expressions above as the product of two other elliptic $\Gamma$ functions with transformed parameters. Applying this identity to the integrand $\Delta_{\mathbb{S}^3\times \mathbb{S}^1}(\ul z)$ and recalling the parametrization (\ref{q_tq_s})
\be
q_\tau=\e^{2\pi\i\frac{\omega}{\omega_1}}~,\quad q_\sigma=\e^{-2\pi\i\frac{\omega_3}{\omega_1}}~,
\ee
we finally arrive at the desired expression
\be\label{S3S1P3}
\boxed{\Delta_{\mathbb{S}^3\times \mathbb{S}^1}(\ul z)=\e^{-\i\pi\mathcal{P}_3(\ul X)}\Delta_{S'}(\ul X)}~.
\ee
The complete parameter identification between $\Delta_{\mathbb{S}^3\times \mathbb{S}^1}(\ul z)$ and $\Delta_{S'}(\ul X)$ is also straightforward and closely follows tables (\ref{gauge:algebra1}), (\ref{gauge:algebra2})
\be\label{gauge:algebra3}
\begin{array}{c|c|c|c|c|c|c|c|c|c}
\textrm{Gauge theory}&\q&\p&\hat T&X_{a,j}&M_{e}&M_e-\bar M_{e}&\bar M_{a,f}&\bar M_{a,f}- M_{a,f}&\kappa_a/\omega_3~\\
\hline
\textrm{$W_{q,t;q'}^{S'}(\Gamma)$}&\e^{2\pi\frac{\omega_1}{\omega_3}}&\e^{2\pi\frac{\omega_2}{\omega_3}}&\beta\omega&X_{a,j}&M_{e}&\omega(1-\beta)& \chi_{a,f}&U_{a,f}&\hat\alpha_a\omega/\omega_1\omega_2~\end{array}\quad~.
\ee

The only things that are left to discuss are the $q$-constants (\ref{qConst}) and, more importantly, the total cubic polynomial in (\ref{S3S1P3}). While $q$-constants can again be pulled out of integration by using the modular property (\ref{ThetaS}) of $\Theta$ functions,\footnote{In fact, $q$-constants are ambiguous in block integrals but disappear on the compact space.} the discussion of the polynomial terms is more delicate and it is related to the obstruction to the complete factorizability of the compact space partition function due to mixed-gauge anomalies. Let us denote by $\#_a$ the total number of loop arrows $e\in\Gamma_1$ at $a\in\Gamma_0$ and by  $\#_{a,b}$ the total number of arrows $e\in\Gamma_1$ connecting $a\in\Gamma_0$ to a different $b\in\Gamma_0$. Then the various contributions to the cubic polynomial attached to the quiver $\Gamma$ are (up background terms encoding global anomalies):
\begin{itemize}
\item \textbf{Node at $a$:}
\be
\mathcal{P}_3(\ul X_{a})=\frac{2(1-\#_a)\hat T}{\omega_1\omega_2\omega_3}(N_a\sum_{j} X_{a,j}^2-(\sum_j X_{a,j})^2)~.
\ee
\item \textbf{Arrows between $a$ and $b$:}
\begin{multline}
\mathcal{P}_3(\ul X_{a},\ul X_{b})=-\frac{(\#_{a,b}+\#_{b,a})\hat T}{\omega_1\omega_2\omega_3}(N_b\sum_j X^2_{a,j}+N_a\sum_k X^2_{b,k}-2\sum_j X_{a,j}\sum_k X_{b,k})+\\
+\frac{\hat T}{\omega_1\omega_2\omega_3}(\#_{a, b}(\hat T-\omega+\omega_3)+2 \sum_{e:a\to b}M_e)(N_b\sum_j X_{a,j}-N_a\sum_k X_{b,k})+\\
-\frac{\hat T}{\omega_1\omega_2\omega_3}(\#_{b, a}(\hat T-\omega-\omega_3)+2\sum_{e:b\to a}M_e)(N_b\sum_j X_{a,j}-N_a\sum_k X_{b,k})~.
\end{multline}
\item \textbf{Fundamentals at $a$:}
\begin{multline}
\mathcal{P}_3(\ul X_{a})=\frac{\sum_f (M_{a,f}-\omega)-\sum_f  \bar M_{a,f}}{\omega_1\omega_2\omega_3}\sum_j X^2_{a,j}+\\
-\frac{(\sum_f(M_{a,f}-\omega)-\sum_f  \bar M_{a,f})(\sum_f (M_{a,f}-\omega_3)+ \bar M_{a,f})}{\omega_1\omega_2\omega_3}\sum_j X_{a,j}~.
\end{multline}
\end{itemize}
These polynomials encode the various mixed-gauge anomalies and must vanish in order to have a sensible quantum theory. Since 
\be
\#_a=1-\frac{C_{aa}^{[0]}}{2}~,\quad \#_{a,b}+\#_{b,a}=- C_{ab}^{[0]}~,\quad a\neq b~,
\ee
by looking at the bi-fundamental contribution for generic values of the parameters we conclude that we must have either
\be
\sum_j X_{a,j}=0~, \quad \textrm{ or }\quad ~C_{ab}^{[0]}=0~,\quad a\neq b~.
\ee
These conditions are met by unitary disconnected quivers and special unitary quivers.

\subsubsection*{Unitary gauge groups.} In this case we can focus on the single node $G={\rm U}(N_a)$ theory. Further restrictions come from the vanishing of the quadratic terms
\be
\hat T C_{aa}^{[0]}(N_a\sum_j X_{a,j}^2-(\sum_j X_{a,j})^2)+(\sum_f( M_{a,f}-\omega)-\sum_f \bar M_{a,f})\sum_j X^2_{a,j}=0~,
\ee
which requires
\be
\#_a=1~,\quad \sum_f(M_{a,f}-\omega)-\sum_f \bar M_{a,f}=0~.
\ee
This example is therefore associated to the elliptic algebra with quiver $\Gamma=\hat A_1$. 

If the theory is $\mathcal{N}=1$ with an R-symmetry we can parametrize the fugacities according to
\be
M_{a,f}=\frac{\omega}{2}R+A_{a,f}~,\quad \bar M_{a,f}=-\frac{\omega}{2}R-\bar A_{a,f}~,
\ee
where we assigned R-charge $R$ to all the chiral multiplets and denoted by $A_{a,f},\bar A_{a,f}$ additional flavor fugacities, with $f=1,\ldots, N_{a,\rm f}$. After imposing the traceless condition $\sum_{f} (A_{a,f}+\bar A_{a,f})=0$ we get the R-charge assignment $R=1$. The theory with no fundamentals has the matter content of the $\mathcal{N}=2^*$ theory.

\subsubsection*{Special unitary groups.}
The traceless condition $\sum_j X_{a,j}=0$ simplifies the total contribution to node $a$ imposing the balancing condition
\be
\hat{T}\sum_b C_{ab}^{[0]}N_b+\sum_f( M_{a,f}-\omega)-\sum_f \bar M_{a,f}=0~.
\ee
For $\mathcal{N}=1$ theories with an R-symmetry we can parametrize the fugacities as in the previous example and impose the traceless condition $\sum_{f} (A_{a,f}+\bar A_{a,f})=0$. A particularly interesting example is provided by the SQCD, in which case we have a single node ${\rm SU}(N_a)$ gauge group without any loop arrow, namely $\Gamma=A_1$, and without the adjoint chiral in the would-be $\mathcal{N}=2$ vector multiplet. As noted in \cite{Gadde:2010en} (see also \cite{Chen:2014rca}), one can integrate out the adjoint chiral by turning on a superpotential mass term, corresponding to the specialization $\hat T=\omega/2$ in the index. In this case one gets the R-charge assignment 
\be
R=\frac{N_{a,\rm f}-N_{a}}{N_{a,\rm f}}~,
\ee
which in the conformal case $2 N_a=N_{a,\rm f}$ (see below) correctly yields $R=1/2$.

For honest $\mathcal{N}=2$ theories the pairs of fundamental and anti-fundamental chiral multiplets must form actual hypers, and hence we have to specialize the fugacities according to 
\be
M_{a,f}=\frac{\omega}{2}-\frac{\hat T}{2}+A_{a,f}~,\quad \bar M_{a,f}=-\frac{\omega}{2}+\frac{\hat T}{2}+A_{a,f}~,
\ee
recovering the anomaly free condition for the (conformal) quiver $\Gamma$ \cite{Nekrasov:2013xda}
\be
\sum_{b}C_{ab}^{[0]}N_b=N_{a,\rm f}~.
\ee

\section{Summary and conclusions}\label{sec:concl}
In this paper we have studied the modular double version of the elliptic $W_{q,t;q'}(\Gamma)$ algebras of \cite{Kimura:2016dys} and their appearance in 4d supersymmetric gauge theories. These algebras were introduced to describe, in the spirit of the BPS/CFT correspondence, $\mathbb{R}^4\times\mathbb{T}^2$ Nekrasov partition functions of supersymmetric $\Gamma$ quiver unitary gauge theories as $W_{q,t;q'}(\Gamma)$ correlators involving infinitely many screening currents. In this paper we have instead focused on the case when there are only finitely many insertions, in which case the most generic observable (the generating function) has the form of an ordinary matrix model. The very same type of matrix model arises when computing for instance the $\mathbb{S}^3\times \mathbb{S}^1$ partition function (index) of 4d supersymmetric gauge theories, and hence we expect that elliptic $W_{q,t;q'}(\Gamma)$ algebras will be useful for studying those theories and the parent 6d theories on compact spaces such as $\mathbb{S}^5\times \mathbb{S}^1$ or $\mathbb{S}^4\times \mathbb{T}^2$ \cite{Lockhart:2012vp,Iqbal:2012xm,Kim:2012qf,Kim:2012ava}, the 4d theories being particular codimension 2 defects. The $\mathbb{S}^5$ case is currently under investigation. A particularly interesting setup which may be studied through $W_{q,t;q'}(\Gamma)$ techniques is provided by the theories arising from doubly compactified toric Calabi-Yau 3-folds \cite{Hollowood:2003cv,Hohenegger:2013ala,Hohenegger:2015btj}, whose partition functions may correspond to torus correlators of elliptic $W_{q,t;q'}(A_n)$ algebras (see appendix \ref{sec:torus} for further comments, torus correlators of $W_{q,t}(\Gamma)$ algebras and the relation with the present work). Another direction worth studying is the connection of the present work with the index of class  $\mathcal{S}$ theories, which is known to have a dual 2d TQFT description  \cite{Gadde:2009kb}. Since we can access at least a subset of those theories using elliptic $W_{q,t,q'}(\Gamma)$ algebra techniques,  we expect to find interesting relations between the elliptic modular double, the 2d TQFT the and related elliptic integrable systems \cite{Gaiotto:2012xa}.

\acknowledgments
We thank G. Festuccia, C. Koz\c{c}az and S. Shakirov for discussions. The research of R.L., F.N. and  M.Z. is supported in part by Vetenskapsr\r{a}det under grant \#2014-5517, by the STINT grant and by the grant ``Geometry and Physics" from the Knut and Alice Wallenberg foundation.

\appendix

\section{Special functions}\label{sec:specialf}
We summarize the special functions and their properties used throughout the paper, for details we refer to \cite{2003math......6164N} and to \cite{Spiridonov:2013zma} for a review with applications to gauge theories and integrable systems. The (multiple) $q$-Pochhammer symbol is defined by
\be\label{qPoch}
(x;q_1,\ldots,q_N)_\infty=\exp\left(-\sum_{n>0}\frac{x^n}{n\prod_{k=1}^N(1-q_k^n)}\right)=\prod_{n_1,\ldots,n_N\geq 0}(1-x q_1^{n_1}\cdots q_N^{n_N})~.
\ee
In the last expression $|q_k|<1$ is assumed, but it can be extended to $|q_k|>1$ by replacing 
\be\label{qPoch:an}
(x;q)_\infty\to \frac{1}{(q^{-1}x;q^{-1})_\infty}~.
\ee

The $\Theta$ function is defined by
\be\label{Theta}
\Theta(x;q)=(x;q)_\infty(q x^{-1};q)_\infty~.
\ee
The modular properties we are interested in are
\be\label{ThetaS}
\Theta(\e^{\frac{2\pi\i}{\omega_1} X};\e^{2\pi\i\frac{\omega}{\omega_1}})\Theta(\e^{\frac{2\pi\i}{\omega_2} X};\e^{2\pi\i\frac{\omega}{\omega_2}})=
\e^{-\i\pi B_{22}(X|\ul\omega)}~,
\ee
and more generally
\be\label{ThetaSr}
\Theta(\e^{2\pi\i X}\e^{\frac{2\pi\i\ell}{r}};\e^{2\pi\i\epsilon})\Theta(\e^{\frac{2\pi\i X}{r\epsilon-1}}\e^{-\frac{2\pi\i\ell}{r}};\e^{\frac{2\pi\i\epsilon}{r\epsilon-1}})=\e^{\frac{\i\pi}{r}\ell(r-\ell)}\e^{-\i\pi \left(B_{22}(X|1,\epsilon)+B_{22}(1+\frac{X}{r\epsilon-1}|1,\frac{\epsilon}{r\epsilon-1})\right)}~,~~~~~
\ee
for $r\in\mathbb{Z}$, $\ell\in\mathbb{Z}_r$, and
\be\label{ThetaA}
\Theta(q^{-\frac{\ell}{2}}x;q)\Theta(q^{\frac{\ell}{2}}x;q^{-1})=(-x)^{\ell+1}~,
\ee
for $\ell\in\mathbb{Z}$. Here $B_{22}(X|\ul\omega)$ is the quadratic Bernoulli polynomial
\be
B_{22}(X|\ul\omega)=\frac{1}{\omega_1\omega_2}\left(\left(X-\frac{\omega}{2}\right)^2-\frac{\omega_1^2+\omega_2^2}{12}\right)~,\quad \omega=\omega_1+\omega_2~.
\ee

The elliptic $\Gamma$ function is defined by
\be
\Gamma(x;p,q)=\frac{(pq x^{-1};p,q)_\infty}{(x;p,q)_\infty}~.
\ee
The elliptic $\Gamma$ function has a very non-trivial behaviour under  modular transformations  \cite{Felder200044}  (${\rm Im}(\omega_j/\omega_k)\neq 0$ is assumed)
\be\label{gammamod}
\Gamma(\e^{\frac{2\pi\i }{\omega_1}X};\e^{2\pi\i\frac{\omega_2}{\omega_1}},\e^{2\pi\i\frac{\omega_3}{\omega_1}})
\Gamma(\e^{\frac{2\pi\i }{\omega_2}X};\e^{2\pi\i\frac{\omega_1}{\omega_2}},\e^{2\pi\i\frac{\omega_3}{\omega_2}})
\Gamma(\e^{\frac{2\pi\i }{\omega_3}X};\e^{2\pi\i\frac{\omega_1}{\omega_3}},\e^{2\pi\i\frac{\omega_2}{\omega_3}})
=\e^{-\frac{\i\pi}{3}B_{33}(X|\ul\omega)}~,
\ee
or
\be\label{gammamod}
\Gamma(\e^{\frac{2\pi\i }{\omega_3}X}\!;\!\e^{2\pi\i\frac{\omega_1}{\omega_3}},\e^{2\pi\i\frac{\omega_2}{\omega_3}})\!=\!\e^{\frac{\i\pi}{3}B_{33}(X|\omega_1,\omega_2,-\omega_3)}\Gamma(\e^{\frac{2\pi\i }{\omega_1}X}\!;\!\e^{2\pi\i\frac{\omega_2}{\omega_1}},\e^{-2\pi\i\frac{\omega_3}{\omega_1}})
\Gamma(\e^{\frac{2\pi\i }{\omega_2}X}\!;\!\e^{2\pi\i\frac{\omega_1}{\omega_2}},\e^{-2\pi\i\frac{\omega_3}{\omega_2}})~,
\ee
where  $B_{33}(X|\ul\omega)$ is the cubic Bernoulli polynomial
\be\label{B33}
B_{33}(X|\ul\omega)=\frac{1}{\omega_1\omega_2\omega_3}\left(X-\frac{\omega}{2}-\frac{\omega_3}{2}\right)\left(\left(X-\frac{\omega}{2}\right)^2-\omega_3\left(X-\frac{\omega}{2}\right)-\frac{\omega_1^2+\omega_2^2}{4}\right)~.
\ee

\section{Torus correlators of trigonometric $W_{q,t}$ algebras}\label{sec:torus}
Elliptic $W_{q,t;q'}(\Gamma)$ algebras have a smooth limit to trigonometric $W_{q,t}(\Gamma)$ algebras \cite{Kimura:2015rgi} when the elliptic parameter $q'$ goes to zero. In analogy with the elliptic case that we have discussed in this work, the generators of $W_{q,t}(\Gamma)$ can be defined as the commutant (up to total differences) of the trigonometric screening currents 
\be
\S^a_{\rm trig}(w)=\; :\e^{\sum_{n\neq 0}\s_{a,n}w^{-n}}:w^{\s_{a,0}}\e^{\tilde{\s}_{a,0}}
\ee
in the trigonometric Heisenberg algebra generated by 
\be\label{Heistrig}
[\s_{a,n},\s_{b,m}]=-\frac{1-t^{-n}}{n(1-q^{-n})}C_{ab}^{[ n]}\delta_{n+m,0}~,\quad n,m\in\mathbb{Z}\backslash\{0\}~,\quad  [\s_{a,0},\tilde{\s}_{b,0}]=\beta C_{ab}^{[0]} ~.
\ee
As we have mentioned in the introduction, to any trigonometric $W_{q,t}(\Gamma)$ algebra one can associate a 3d $\mathcal{N}=2$ unitary quiver gauge theory and there is a dictionary between observables on the two sides. This is the trigonometric/3d limit of what we have discussed in the present paper. So far, most of the existing examples are given for {\it sphere correlators} of $W_{q,t}(\Gamma)$ vertex operators on the one hand, and gauge theory partition functions on 3-manifolds $\mathcal{M}_3$ on the other hand. Interestingly, {\it torus correlators} in trigonometric $W_{q,t}(\Gamma)$ algebras correspond to sphere correlators in elliptic $W_{q,t;q'}(\Gamma)$ algebras, and hence observables in the lifted gauge theory on $\mathcal{M}_3\times S^1$. The trace operation in the torus correlator is associated with the additional tower of KK modes. The computation of the $\mathbb{S}^3\times \mathbb{S}^1$ partition function of 4d $\mathcal{N}=1$ gauge theories from a tower of 3d theories on $\mathbb{S}^3$ was considered in \cite{Assel:2014paa} (see also \cite{DiPietro:2014bca,DiPietro:2016ond,Ardehali:2015bla,Ardehali:2015hya}). In this section we will discuss the algebraic side of this relation. For simplicity we will work with standard $W_{q,t}(\Gamma)$ algebras, but the results will go through the modular double construction. Let us introduce a grading operator $\L_0$ which satisfies
\be
[\L_0,\s_{a,n}]=-n \s_{a,n}~,
\ee
and let us try to compute the torus correlator
\be
\chi_{\ul\alpha}(\O)=\textrm{Tr}_{\mathcal{F}_{\ul \alpha}}\left(q'^{\L_0}\O\right)~,
\ee
where  $\O$ is some operator written in terms of the non-zero modes of the Heisenberg algebra. The trace is taken over the entire Fock space. The Clavelli-Shapiro trace technique \cite{Clavelli:1973uk} tells us that we can equivalently compute (up to a $q'$-dependent constant)
\be
\chi_{\ul\alpha}(\O)=\bra{\ul\alpha}\hat{\O}\ket{\ul\alpha}~,
\ee
where $\hat{\O}$ is the elliptic version of the operator $\O$, namely the former is obtained from the latter through the substitution 
\be
\s_{a,n}\to \frac{\a_{a,n}}{1-q'^n}+\b_{a,-n}~,\quad \s_{a,-n}\to \a_{a,-n}-\frac{\b_{a,n}}{1-q'^{-n}}~,\quad n>0~,
\ee
where the oscillators $\a_{a,n},\b_{a,n}$ satisfy the trigonometric Heisenberg algebra (\ref{Heistrig}) and commute with each other. We can now identify the oscillators
\be
\s_{a,n}^{(+)}=\frac{\a_{a,n}}{1-q'^n}~,\quad s_{a,-n}^{(+)}=\a_{a,-n}~,\quad \s_{a,n}^{(-)}=-\frac{\b_{a,n}}{1-q'^{-n}}~,\quad s_{a,-n}^{(-)}=\b_{a,-n}~,\quad n>0~,
\ee
which indeed satisfy the elliptic Heisenberg algebra (\ref{Heisell}), and write the desired result
\be
\bra{\ul\alpha_\infty}\; \cdots\;\Z\ket{\ul\alpha}=\textrm{Tr}_{\mathcal{F}_{\ul\alpha}}\left(\e^{\ln q'\L_0-\sum_a \tilde\s_{0,a}N_a}\;\cdots\;\oint\frac{\rd \ul w}{2\pi\i\ul w}\prod_{a=1}^{|\Gamma_0|}\prod_{j=1}^{N_a}\S_{\rm trig}^a(w_{a,j})\right)~,
\ee
where $(\alpha_{\infty})_a=\alpha_a+\beta C_{ab}^{[0]}N_b$ to ensure momentum conservation and the dots represent the insertion of additional vertex operators of the type (\ref{Vertex}).

This kind of relation between trigonometric and elliptic algebras is also related to the fiber/base duality in the context of 5d or 6d theories \cite{Mironov:2016cyq} arising from toric Calabi-Yau 3-folds with a periodic direction. In one case the periodic direction is identified with the presence of additional adjoint matter, in the dual picture it is identified with an additional space-time direction. It is worth noting that this relation does not imply that all the elliptic $W_{q,t;q'}(\Gamma)$ algebra observables can be recast in terms of the $W_{q,t}(\Gamma)$ algebra ones, for instance operators involving $\s_{n,a}^{(+)}$ and $\s_{n,a}^{(-)}$ asymmetrically, or torus correlators in elliptic $W_{q,t;q'}(\Gamma)$ algebras. Moreover, the fact that the elliptic deformation leads to a well-defined associative algebra seems to be a non-trivial fact.

\bibliographystyle{JHEP}
\bibliography{ell_mod_double_refs}

\providecommand{\href}[2]{#2}\begingroup\raggedright\begin{thebibliography}{100}

\bibitem{Nekrasov:2015wsu}
N.~Nekrasov, \emph{Bps/cft correspondence: non-perturbative dyson-schwinger
  equations and qq-characters},
  \href{http://dx.doi.org/10.1007/JHEP03(2016)181}{\emph{JHEP} {\bf 03} (2016)
  181}, [\href{http://arxiv.org/abs/1512.05388}{{\tt 1512.05388}}].

\bibitem{Nekrasov:2016qym}
N.~Nekrasov, \emph{{BPS/CFT correspondence II: Instantons at crossroads, Moduli
  and Compactness Theorem}},  \href{http://arxiv.org/abs/1608.07272}{{\tt
  1608.07272}}.

\bibitem{Nekrasov:2016ydq}
N.~Nekrasov, \emph{{BPS/CFT Correspondence III: Gauge Origami partition
  function and qq-characters}},  \href{http://arxiv.org/abs/1701.00189}{{\tt
  1701.00189}}.

\bibitem{Nekrasov:2009uh}
N.~A. Nekrasov and S.~L. Shatashvili, \emph{{Supersymmetric vacua and Bethe
  ansatz}},
  \href{http://dx.doi.org/10.1016/j.nuclphysbps.2009.07.047}{\emph{Nucl. Phys.
  Proc. Suppl.} {\bf 192-193} (2009) 91--112},
  [\href{http://arxiv.org/abs/0901.4744}{{\tt 0901.4744}}].

\bibitem{Nekrasov:2009ui}
N.~A. Nekrasov and S.~L. Shatashvili, \emph{{Quantum integrability and
  supersymmetric vacua}},
  \href{http://dx.doi.org/10.1143/PTPS.177.105}{\emph{Prog. Theor. Phys.
  Suppl.} {\bf 177} (2009) 105--119},
  [\href{http://arxiv.org/abs/0901.4748}{{\tt 0901.4748}}].

\bibitem{Nekrasov:2009rc}
N.~A. Nekrasov and S.~L. Shatashvili, \emph{{Quantization of Integrable Systems
  and Four Dimensional Gauge Theories}},  in \emph{{Proceedings, 16th
  International Congress on Mathematical Physics (ICMP09)}}, 2009.
\newblock \href{http://arxiv.org/abs/0908.4052}{{\tt 0908.4052}}.

\bibitem{Beem:2013sza}
C.~Beem, M.~Lemos, P.~Liendo, W.~Peelaers, L.~Rastelli and B.~C. van Rees,
  \emph{{Infinite Chiral Symmetry in Four Dimensions}},
  \href{http://dx.doi.org/10.1007/s00220-014-2272-x}{\emph{Commun. Math. Phys.}
  {\bf 336} (2015) 1359--1433}, [\href{http://arxiv.org/abs/1312.5344}{{\tt
  1312.5344}}].

\bibitem{Beem:2014kka}
C.~Beem, L.~Rastelli and B.~C. van Rees, \emph{{$ \mathcal{W} $ symmetry in six
  dimensions}}, \href{http://dx.doi.org/10.1007/JHEP05(2015)017}{\emph{JHEP}
  {\bf 05} (2015) 017}, [\href{http://arxiv.org/abs/1404.1079}{{\tt
  1404.1079}}].

\bibitem{Alday:2009aq}
L.~F. Alday, D.~Gaiotto and Y.~Tachikawa, \emph{{Liouville Correlation
  Functions from Four-dimensional Gauge Theories}},
  \href{http://dx.doi.org/10.1007/s11005-010-0369-5}{\emph{Lett. Math. Phys.}
  {\bf 91} (2010) 167--197}, [\href{http://arxiv.org/abs/0906.3219}{{\tt
  0906.3219}}].

\bibitem{Wyllard:2009hg}
N.~Wyllard, \emph{A(n-1) conformal toda field theory correlation functions from
  conformal n = 2 su(n) quiver gauge theories},
  \href{http://dx.doi.org/10.1088/1126-6708/2009/11/002}{\emph{JHEP} {\bf 11}
  (2009) 002}, [\href{http://arxiv.org/abs/0907.2189}{{\tt 0907.2189}}].

\bibitem{Pestun:2007rz}
V.~Pestun, \emph{{Localization of gauge theory on a four-sphere and
  supersymmetric Wilson loops}},
  \href{http://dx.doi.org/10.1007/s00220-012-1485-0}{\emph{Commun. Math. Phys.}
  {\bf 313} (2012) 71--129}, [\href{http://arxiv.org/abs/0712.2824}{{\tt
  0712.2824}}].

\bibitem{Hama:2012bg}
N.~Hama and K.~Hosomichi, \emph{{Seiberg-Witten Theories on Ellipsoids}},
  \href{http://dx.doi.org/10.1007/JHEP09(2012)033,
  10.1007/JHEP10(2012)051}{\emph{JHEP} {\bf 09} (2012) 033},
  [\href{http://arxiv.org/abs/1206.6359}{{\tt 1206.6359}}].

\bibitem{Gaiotto:2009we}
D.~Gaiotto, \emph{{N=2 dualities}},
  \href{http://dx.doi.org/10.1007/JHEP08(2012)034}{\emph{JHEP} {\bf 08} (2012)
  034}, [\href{http://arxiv.org/abs/0904.2715}{{\tt 0904.2715}}].

\bibitem{Nekrasov:2002qd}
N.~A. Nekrasov, \emph{{Seiberg-Witten prepotential from instanton counting}},
  \href{http://dx.doi.org/10.4310/ATMP.2003.v7.n5.a4}{\emph{Adv. Theor. Math.
  Phys.} {\bf 7} (2003) 831--864},
  [\href{http://arxiv.org/abs/hep-th/0206161}{{\tt hep-th/0206161}}].

\bibitem{Nekrasov:2003rj}
N.~Nekrasov and A.~Okounkov, \emph{{Seiberg-Witten theory and random
  partitions}}, \href{http://dx.doi.org/10.1007/0-8176-4467-9_15}{\emph{Prog.
  Math.} {\bf 244} (2006) 525--596},
  [\href{http://arxiv.org/abs/hep-th/0306238}{{\tt hep-th/0306238}}].

\bibitem{Teschner:2016yzf}
J.~Teschner, ed., \emph{{New Dualities of Supersymmetric Gauge Theories}}.
\newblock Mathematical Physics Studies. Springer, Cham, Switzerland, 2016,
  \href{http://dx.doi.org/10.1007/978-3-319-18769-3}{10.1007/978-3-319-18769-3}.

\bibitem{Awata:2009ur}
H.~Awata and Y.~Yamada, \emph{{Five-dimensional AGT Conjecture and the Deformed
  Virasoro Algebra}},
  \href{http://dx.doi.org/10.1007/JHEP01(2010)125}{\emph{JHEP} {\bf 01} (2010)
  125}, [\href{http://arxiv.org/abs/0910.4431}{{\tt 0910.4431}}].

\bibitem{Awata:2010yy}
H.~Awata and Y.~Yamada, \emph{{Five-dimensional AGT Relation and the Deformed
  beta-ensemble}}, \href{http://dx.doi.org/10.1143/PTP.124.227}{\emph{Prog.
  Theor. Phys.} {\bf 124} (2010) 227--262},
  [\href{http://arxiv.org/abs/1004.5122}{{\tt 1004.5122}}].

\bibitem{Awata:2011dc}
H.~Awata, B.~Feigin, A.~Hoshino, M.~Kanai, J.~Shiraishi and S.~Yanagida,
  \emph{{Notes on Ding-Iohara algebra and AGT conjecture}},
  \href{http://arxiv.org/abs/1106.4088}{{\tt 1106.4088}}.

\bibitem{Mironov:2011dk}
A.~Mironov, A.~Morozov, S.~Shakirov and A.~Smirnov, \emph{{Proving AGT
  conjecture as HS duality: extension to five dimensions}},
  \href{http://dx.doi.org/10.1016/j.nuclphysb.2011.09.021}{\emph{Nucl. Phys.}
  {\bf B855} (2012) 128--151}, [\href{http://arxiv.org/abs/1105.0948}{{\tt
  1105.0948}}].

\bibitem{Awata:2016riz}
H.~Awata, H.~Kanno, T.~Matsumoto, A.~Mironov, A.~Morozov, A.~Morozov et~al.,
  \emph{{Explicit examples of DIM constraints for network matrix models}},
  \href{http://arxiv.org/abs/1604.08366}{{\tt 1604.08366}}.

\bibitem{Mironov:2016yue}
A.~Mironov, A.~Morozov and Y.~Zenkevich, \emph{{Ding-Iohara-Miki symmetry of
  network matrix models}},  \href{http://arxiv.org/abs/1603.05467}{{\tt
  1603.05467}}.

\bibitem{Awata:2016bdm}
H.~Awata, H.~Kanno, A.~Mironov, A.~Morozov, A.~Morozov, Y.~Ohkubo et~al.,
  \emph{{Anomaly in RTT relation for DIM algebra and network matrix models}},
  \href{http://arxiv.org/abs/1611.07304}{{\tt 1611.07304}}.

\bibitem{Kimura:2015rgi}
T.~Kimura and V.~Pestun, \emph{{Quiver W-algebras}},
  \href{http://arxiv.org/abs/1512.08533}{{\tt 1512.08533}}.

\bibitem{Nieri:2013yra}
F.~Nieri, S.~Pasquetti and F.~Passerini, \emph{{3d and 5d Gauge Theory
  Partition Functions as $q$-deformed CFT Correlators}},
  \href{http://dx.doi.org/10.1007/s11005-014-0727-9}{\emph{Lett. Math. Phys.}
  {\bf 105} (2015) 109--148}, [\href{http://arxiv.org/abs/1303.2626}{{\tt
  1303.2626}}].

\bibitem{Nieri:2013vba}
F.~Nieri, S.~Pasquetti, F.~Passerini and A.~Torrielli, \emph{5d partition
  functions, q-virasoro systems and integrable spin-chains},
  \href{http://dx.doi.org/10.1007/JHEP12(2014)040}{\emph{JHEP} {\bf 12} (2014)
  040}, [\href{http://arxiv.org/abs/1312.1294}{{\tt 1312.1294}}].

\bibitem{Carlsson:2013jka}
E.~Carlsson, N.~Nekrasov and A.~Okounkov, \emph{{Five dimensional gauge
  theories and vertex operators}},  \href{http://arxiv.org/abs/1308.2465}{{\tt
  1308.2465}}.

\bibitem{Shiraishi:1995rp}
J.~Shiraishi, H.~Kubo, H.~Awata and S.~Odake, \emph{A quantum deformation of
  the virasoro algebra and the macdonald symmetric functions},
  \href{http://dx.doi.org/10.1007/BF00398297}{\emph{Lett. Math. Phys.} {\bf 38}
  (1996) 33--51}, [\href{http://arxiv.org/abs/q-alg/9507034}{{\tt
  q-alg/9507034}}].

\bibitem{Awata:1995zk}
H.~Awata, H.~Kubo, S.~Odake and J.~Shiraishi, \emph{{Quantum W(N) algebras and
  Macdonald polynomials}},
  \href{http://dx.doi.org/10.1007/BF02102595}{\emph{Commun. Math. Phys.} {\bf
  179} (1996) 401--416}, [\href{http://arxiv.org/abs/q-alg/9508011}{{\tt
  q-alg/9508011}}].

\bibitem{1997q.alg.....8006F}
E.~{Frenkel} and N.~{Reshetikhin}, \emph{{Deformations of W-algebras associated
  to simple Lie algebras}},  in \emph{eprint arXiv:q-alg/9708006}, Aug., 1997.

\bibitem{Aganagic:2013tta}
M.~Aganagic, N.~Haouzi, C.~Kozcaz and S.~Shakirov, \emph{{Gauge/Liouville
  Triality}},  \href{http://arxiv.org/abs/1309.1687}{{\tt 1309.1687}}.

\bibitem{Aganagic:2014oia}
M.~Aganagic, N.~Haouzi and S.~Shakirov, \emph{{$A_n$-Triality}},
  \href{http://arxiv.org/abs/1403.3657}{{\tt 1403.3657}}.

\bibitem{Nedelin:2016gwu}
A.~Nedelin, F.~Nieri and M.~Zabzine, \emph{{$q$-Virasoro modular double and 3d
  partition functions}},  \href{http://arxiv.org/abs/1605.07029}{{\tt
  1605.07029}}.

\bibitem{Pasquetti:2011fj}
S.~Pasquetti, \emph{{Factorisation of N = 2 Theories on the Squashed
  3-Sphere}}, \href{http://dx.doi.org/10.1007/JHEP04(2012)120}{\emph{JHEP} {\bf
  04} (2012) 120}, [\href{http://arxiv.org/abs/1111.6905}{{\tt 1111.6905}}].

\bibitem{Beem:2012mb}
C.~Beem, T.~Dimofte and S.~Pasquetti, \emph{{Holomorphic Blocks in Three
  Dimensions}}, \href{http://dx.doi.org/10.1007/JHEP12(2014)177}{\emph{JHEP}
  {\bf 12} (2014) 177}, [\href{http://arxiv.org/abs/1211.1986}{{\tt
  1211.1986}}].

\bibitem{Nieri:2015dts}
F.~Nieri, \emph{{An elliptic Virasoro symmetry in 6d}},
  \href{http://arxiv.org/abs/1511.00574}{{\tt 1511.00574}}.

\bibitem{Iqbal:2015fvd}
A.~Iqbal, C.~Kozcaz and S.-T. Yau, \emph{{Elliptic Virasoro Conformal Blocks}},
   \href{http://arxiv.org/abs/1511.00458}{{\tt 1511.00458}}.

\bibitem{Kimura:2016dys}
T.~Kimura and V.~Pestun, \emph{{Quiver elliptic W-algebras}},
  \href{http://arxiv.org/abs/1608.04651}{{\tt 1608.04651}}.

\bibitem{Tan:2016cky}
M.-C. Tan, \emph{{Higher AGT Correspondences, W-algebras, and Higher Quantum
  Geometric Langlands Duality from M-Theory}},
  \href{http://arxiv.org/abs/1607.08330}{{\tt 1607.08330}}.

\bibitem{Bazhanov:2010kz}
V.~V. Bazhanov and S.~M. Sergeev, \emph{{A Master solution of the quantum
  Yang-Baxter equation and classical discrete integrable equations}},
  \href{http://dx.doi.org/10.4310/ATMP.2012.v16.n1.a3}{\emph{Adv. Theor. Math.
  Phys.} {\bf 16} (2012) 65--95}, [\href{http://arxiv.org/abs/1006.0651}{{\tt
  1006.0651}}].

\bibitem{Bazhanov:2011mz}
V.~V. Bazhanov and S.~M. Sergeev, \emph{{Elliptic gamma-function and multi-spin
  solutions of the Yang-Baxter equation}},
  \href{http://dx.doi.org/10.1016/j.nuclphysb.2011.10.032}{\emph{Nucl. Phys.}
  {\bf B856} (2012) 475--496}, [\href{http://arxiv.org/abs/1106.5874}{{\tt
  1106.5874}}].

\bibitem{Spiridonov:2010em}
V.~P. Spiridonov, \emph{{Elliptic beta integrals and solvable models of
  statistical mechanics}}, {\emph{Contemp. Math.} {\bf 563} (2012) 181--211},
  [\href{http://arxiv.org/abs/1011.3798}{{\tt 1011.3798}}].

\bibitem{Yamazaki:2012cp}
M.~Yamazaki, \emph{{Quivers, YBE and 3-manifolds}},
  \href{http://dx.doi.org/10.1007/JHEP05(2012)147}{\emph{JHEP} {\bf 05} (2012)
  147}, [\href{http://arxiv.org/abs/1203.5784}{{\tt 1203.5784}}].

\bibitem{Yamazaki:2013nra}
M.~Yamazaki, \emph{{New Integrable Models from the Gauge/YBE Correspondence}},
  \href{http://dx.doi.org/10.1007/s10955-013-0884-8}{\emph{J. Statist. Phys.}
  {\bf 154} (2014) 895}, [\href{http://arxiv.org/abs/1307.1128}{{\tt
  1307.1128}}].

\bibitem{Yagi:2015lha}
J.~Yagi, \emph{{Quiver gauge theories and integrable lattice models}},
  \href{http://dx.doi.org/10.1007/JHEP10(2015)065}{\emph{JHEP} {\bf 10} (2015)
  065}, [\href{http://arxiv.org/abs/1504.04055}{{\tt 1504.04055}}].

\bibitem{Maruyoshi:2016caf}
K.~Maruyoshi and J.~Yagi, \emph{{Surface defects as transfer matrices}},
  \href{http://dx.doi.org/10.1093/ptep/ptw151}{\emph{PTEP} {\bf 2016} (2016)
  113B01}, [\href{http://arxiv.org/abs/1606.01041}{{\tt 1606.01041}}].

\bibitem{Yagi:2017hmj}
J.~Yagi, \emph{{Surface defects and elliptic quantum groups}},
  \href{http://arxiv.org/abs/1701.05562}{{\tt 1701.05562}}.

\bibitem{2008arXiv0801.4137S}
V.~P. {Spiridonov}, \emph{{Continuous biorthogonality of the elliptic
  hypergeometric function}}, {\emph{ArXiv e-prints} (Jan., 2008) },
  [\href{http://arxiv.org/abs/0801.4137}{{\tt 0801.4137}}].

\bibitem{Faddeev:1999fe}
L.~D. Faddeev, \emph{{Modular double of quantum group}}, {\emph{Math. Phys.
  Stud.} {\bf 21} (2000) 149--156},
  [\href{http://arxiv.org/abs/math/9912078}{{\tt math/9912078}}].

\bibitem{Ponsot:1999uf}
B.~Ponsot and J.~Teschner, \emph{{Liouville bootstrap via harmonic analysis on
  a noncompact quantum group}},
  \href{http://arxiv.org/abs/hep-th/9911110}{{\tt hep-th/9911110}}.

\bibitem{Yoshida:2014qwa}
Y.~Yoshida, \emph{{Factorization of 4d N=1 superconformal index}},
  \href{http://arxiv.org/abs/1403.0891}{{\tt 1403.0891}}.

\bibitem{Peelaers:2014ima}
W.~Peelaers, \emph{{Higgs branch localization of $ \mathcal{N} $ = 1 theories
  on S$^{3}$ x S$^{1}$}},
  \href{http://dx.doi.org/10.1007/JHEP08(2014)060}{\emph{JHEP} {\bf 08} (2014)
  060}, [\href{http://arxiv.org/abs/1403.2711}{{\tt 1403.2711}}].

\bibitem{Chen:2014rca}
H.-Y. Chen and H.-Y. Chen, \emph{{Heterotic Surface Defects and Dualities from
  2d/4d Indices}}, \href{http://dx.doi.org/10.1007/JHEP10(2014)004}{\emph{JHEP}
  {\bf 10} (2014) 004}, [\href{http://arxiv.org/abs/1407.4587}{{\tt
  1407.4587}}].

\bibitem{Nieri:2015yia}
F.~Nieri and S.~Pasquetti, \emph{{Factorisation and holomorphic blocks in 4d}},
  \href{http://dx.doi.org/10.1007/JHEP11(2015)155}{\emph{JHEP} {\bf 11} (2015)
  155}, [\href{http://arxiv.org/abs/1507.00261}{{\tt 1507.00261}}].

\bibitem{Gadde:2013sca}
A.~Gadde, S.~Gukov and P.~Putrov, \emph{{Fivebranes and 4-manifolds}},
  \href{http://arxiv.org/abs/1306.4320}{{\tt 1306.4320}}.

\bibitem{Festuccia:2011ws}
G.~Festuccia and N.~Seiberg, \emph{{Rigid Supersymmetric Theories in Curved
  Superspace}}, \href{http://dx.doi.org/10.1007/JHEP06(2011)114}{\emph{JHEP}
  {\bf 06} (2011) 114}, [\href{http://arxiv.org/abs/1105.0689}{{\tt
  1105.0689}}].

\bibitem{Dumitrescu:2012ha}
T.~T. Dumitrescu, G.~Festuccia and N.~Seiberg, \emph{{Exploring Curved
  Superspace}}, \href{http://dx.doi.org/10.1007/JHEP08(2012)141}{\emph{JHEP}
  {\bf 08} (2012) 141}, [\href{http://arxiv.org/abs/1205.1115}{{\tt
  1205.1115}}].

\bibitem{Klare:2012gn}
C.~Klare, A.~Tomasiello and A.~Zaffaroni, \emph{{Supersymmetry on Curved Spaces
  and Holography}},
  \href{http://dx.doi.org/10.1007/JHEP08(2012)061}{\emph{JHEP} {\bf 08} (2012)
  061}, [\href{http://arxiv.org/abs/1205.1062}{{\tt 1205.1062}}].

\bibitem{Klare:2013dka}
C.~Klare and A.~Zaffaroni, \emph{{Extended Supersymmetry on Curved Spaces}},
  \href{http://dx.doi.org/10.1007/JHEP10(2013)218}{\emph{JHEP} {\bf 10} (2013)
  218}, [\href{http://arxiv.org/abs/1308.1102}{{\tt 1308.1102}}].

\bibitem{Pestun:2016jze}
V.~Pestun and M.~Zabzine, \emph{{Introduction to localization in quantum field
  theory}},  \href{http://arxiv.org/abs/1608.02953}{{\tt 1608.02953}}.

\bibitem{Assel:2014paa}
B.~Assel, D.~Cassani and D.~Martelli, \emph{{Localization on Hopf surfaces}},
  \href{http://dx.doi.org/10.1007/JHEP08(2014)123}{\emph{JHEP} {\bf 08} (2014)
  123}, [\href{http://arxiv.org/abs/1405.5144}{{\tt 1405.5144}}].

\bibitem{Romelsberger:2005eg}
C.~Romelsberger, \emph{{Counting chiral primaries in N = 1, d=4 superconformal
  field theories}},
  \href{http://dx.doi.org/10.1016/j.nuclphysb.2006.03.037}{\emph{Nucl. Phys.}
  {\bf B747} (2006) 329--353}, [\href{http://arxiv.org/abs/hep-th/0510060}{{\tt
  hep-th/0510060}}].

\bibitem{Kinney:2005ej}
J.~Kinney, J.~M. Maldacena, S.~Minwalla and S.~Raju, \emph{{An Index for 4
  dimensional super conformal theories}},
  \href{http://dx.doi.org/10.1007/s00220-007-0258-7}{\emph{Commun. Math. Phys.}
  {\bf 275} (2007) 209--254}, [\href{http://arxiv.org/abs/hep-th/0510251}{{\tt
  hep-th/0510251}}].

\bibitem{Gadde:2013ftv}
A.~Gadde and S.~Gukov, \emph{{2d Index and Surface operators}},
  \href{http://dx.doi.org/10.1007/JHEP03(2014)080}{\emph{JHEP} {\bf 03} (2014)
  080}, [\href{http://arxiv.org/abs/1305.0266}{{\tt 1305.0266}}].

\bibitem{Benini:2013nda}
F.~Benini, R.~Eager, K.~Hori and Y.~Tachikawa, \emph{{Elliptic genera of
  two-dimensional N=2 gauge theories with rank-one gauge groups}},
  \href{http://dx.doi.org/10.1007/s11005-013-0673-y}{\emph{Lett. Math. Phys.}
  {\bf 104} (2014) 465--493}, [\href{http://arxiv.org/abs/1305.0533}{{\tt
  1305.0533}}].

\bibitem{Benini:2013xpa}
F.~Benini, R.~Eager, K.~Hori and Y.~Tachikawa, \emph{{Elliptic Genera of 2d
  ${\mathcal{N}}$ = 2 Gauge Theories}},
  \href{http://dx.doi.org/10.1007/s00220-014-2210-y}{\emph{Commun. Math. Phys.}
  {\bf 333} (2015) 1241--1286}, [\href{http://arxiv.org/abs/1308.4896}{{\tt
  1308.4896}}].

\bibitem{Gaiotto:2012xa}
D.~Gaiotto, L.~Rastelli and S.~S. Razamat, \emph{{Bootstrapping the
  superconformal index with surface defects}},
  \href{http://dx.doi.org/10.1007/JHEP01(2013)022}{\emph{JHEP} {\bf 01} (2013)
  022}, [\href{http://arxiv.org/abs/1207.3577}{{\tt 1207.3577}}].

\bibitem{Bullimore:2014nla}
M.~Bullimore, M.~Fluder, L.~Hollands and P.~Richmond, \emph{{The superconformal
  index and an elliptic algebra of surface defects}},
  \href{http://dx.doi.org/10.1007/JHEP10(2014)062}{\emph{JHEP} {\bf 10} (2014)
  062}, [\href{http://arxiv.org/abs/1401.3379}{{\tt 1401.3379}}].

\bibitem{Hama:2011ea}
N.~Hama, K.~Hosomichi and S.~Lee, \emph{{SUSY Gauge Theories on Squashed
  Three-Spheres}}, \href{http://dx.doi.org/10.1007/JHEP05(2011)014}{\emph{JHEP}
  {\bf 05} (2011) 014}, [\href{http://arxiv.org/abs/1102.4716}{{\tt
  1102.4716}}].

\bibitem{Imamura:2011wg}
Y.~Imamura and D.~Yokoyama, \emph{{N=2 supersymmetric theories on squashed
  three-sphere}},
  \href{http://dx.doi.org/10.1103/PhysRevD.85.025015}{\emph{Phys. Rev.} {\bf
  D85} (2012) 025015}, [\href{http://arxiv.org/abs/1109.4734}{{\tt
  1109.4734}}].

\bibitem{Alday:2013lba}
L.~F. Alday, D.~Martelli, P.~Richmond and J.~Sparks, \emph{{Localization on
  Three-Manifolds}},
  \href{http://dx.doi.org/10.1007/JHEP10(2013)095}{\emph{JHEP} {\bf 10} (2013)
  095}, [\href{http://arxiv.org/abs/1307.6848}{{\tt 1307.6848}}].

\bibitem{Hosomichi:2014hja}
K.~Hosomichi, \emph{{A review on SUSY gauge theories on $\mathbf S^3$}},  in
  \emph{New Dualities of Supersymmetric Gauge Theories} (J.~Teschner, ed.),
  pp.~307--338.
\newblock 2016.
\newblock \href{http://arxiv.org/abs/1412.7128}{{\tt 1412.7128}}.
\newblock \href{http://dx.doi.org/10.1007/978-3-319-18769-3_10}{DOI}.

\bibitem{Kapustin:2009kz}
A.~Kapustin, B.~Willett and I.~Yaakov, \emph{Exact results for wilson loops in
  superconformal chern-simons theories with matter},
  \href{http://dx.doi.org/10.1007/JHEP03(2010)089}{\emph{JHEP} {\bf 03} (2010)
  089}, [\href{http://arxiv.org/abs/0909.4559}{{\tt 0909.4559}}].

\bibitem{Tanaka:2012nr}
A.~Tanaka, \emph{{Comments on knotted 1/2 BPS Wilson loops}},
  \href{http://dx.doi.org/10.1007/JHEP07(2012)097}{\emph{JHEP} {\bf 07} (2012)
  097}, [\href{http://arxiv.org/abs/1204.5975}{{\tt 1204.5975}}].

\bibitem{Vafa:1994tf}
C.~Vafa and E.~Witten, \emph{{A Strong coupling test of S duality}},
  \href{http://dx.doi.org/10.1016/0550-3213(94)90097-3}{\emph{Nucl. Phys.} {\bf
  B431} (1994) 3--77}, [\href{http://arxiv.org/abs/hep-th/9408074}{{\tt
  hep-th/9408074}}].

\bibitem{Benini:2011nc}
F.~Benini, T.~Nishioka and M.~Yamazaki, \emph{{4d Index to 3d Index and 2d
  TQFT}}, \href{http://dx.doi.org/10.1103/PhysRevD.86.065015}{\emph{Phys. Rev.}
  {\bf D86} (2012) 065015}, [\href{http://arxiv.org/abs/1109.0283}{{\tt
  1109.0283}}].

\bibitem{Razamat:2013jxa}
S.~S. Razamat and M.~Yamazaki, \emph{{S-duality and the N=2 Lens Space Index}},
  \href{http://dx.doi.org/10.1007/JHEP10(2013)048}{\emph{JHEP} {\bf 10} (2013)
  048}, [\href{http://arxiv.org/abs/1306.1543}{{\tt 1306.1543}}].

\bibitem{Nishioka:2014zpa}
T.~Nishioka and I.~Yaakov, \emph{{Generalized indices for $ \mathcal{N} $ = 1
  theories in four-dimensions}},
  \href{http://dx.doi.org/10.1007/JHEP12(2014)150}{\emph{JHEP} {\bf 12} (2014)
  150}, [\href{http://arxiv.org/abs/1407.8520}{{\tt 1407.8520}}].

\bibitem{Closset:2013sxa}
C.~Closset and I.~Shamir, \emph{{The $\mathcal{N}=1$ Chiral Multiplet on
  $T^2\times S^2$ and Supersymmetric Localization}},
  \href{http://dx.doi.org/10.1007/JHEP03(2014)040}{\emph{JHEP} {\bf 03} (2014)
  040}, [\href{http://arxiv.org/abs/1311.2430}{{\tt 1311.2430}}].

\bibitem{Honda:2015yha}
M.~Honda and Y.~Yoshida, \emph{{Supersymmetric index on $T^2 \times S^2$ and
  elliptic genus}},  \href{http://arxiv.org/abs/1504.04355}{{\tt 1504.04355}}.

\bibitem{Benini:2015noa}
F.~Benini and A.~Zaffaroni, \emph{{A topologically twisted index for
  three-dimensional supersymmetric theories}},
  \href{http://dx.doi.org/10.1007/JHEP07(2015)127}{\emph{JHEP} {\bf 07} (2015)
  127}, [\href{http://arxiv.org/abs/1504.03698}{{\tt 1504.03698}}].

\bibitem{Imamura:2011su}
Y.~Imamura and S.~Yokoyama, \emph{Index for three dimensional superconformal
  field theories with general r-charge assignments},
  \href{http://dx.doi.org/10.1007/JHEP04(2011)007}{\emph{JHEP} {\bf 04} (2011)
  007}, [\href{http://arxiv.org/abs/1101.0557}{{\tt 1101.0557}}].

\bibitem{Kapustin:2011jm}
A.~Kapustin and B.~Willett, \emph{{Generalized Superconformal Index for Three
  Dimensional Field Theories}},  \href{http://arxiv.org/abs/1106.2484}{{\tt
  1106.2484}}.

\bibitem{Spiridonov:2011hf}
V.~P. Spiridonov and G.~S. Vartanov, \emph{{Elliptic hypergeometry of
  supersymmetric dualities II. Orthogonal groups, knots, and vortices}},
  \href{http://dx.doi.org/10.1007/s00220-013-1861-4}{\emph{Commun. Math. Phys.}
  {\bf 325} (2014) 421--486}, [\href{http://arxiv.org/abs/1107.5788}{{\tt
  1107.5788}}].

\bibitem{Gadde:2010en}
A.~Gadde, L.~Rastelli, S.~S. Razamat and W.~Yan, \emph{{On the Superconformal
  Index of N=1 IR Fixed Points: A Holographic Check}},
  \href{http://dx.doi.org/10.1007/JHEP03(2011)041}{\emph{JHEP} {\bf 03} (2011)
  041}, [\href{http://arxiv.org/abs/1011.5278}{{\tt 1011.5278}}].

\bibitem{Nekrasov:2013xda}
N.~Nekrasov, V.~Pestun and S.~Shatashvili, \emph{{Quantum geometry and quiver
  gauge theories}},  \href{http://arxiv.org/abs/1312.6689}{{\tt 1312.6689}}.

\bibitem{Lockhart:2012vp}
G.~Lockhart and C.~Vafa, \emph{{Superconformal Partition Functions and
  Non-perturbative Topological Strings}},
  \href{http://arxiv.org/abs/1210.5909}{{\tt 1210.5909}}.

\bibitem{Iqbal:2012xm}
A.~Iqbal and C.~Vafa, \emph{{BPS Degeneracies and Superconformal Index in
  Diverse Dimensions}},
  \href{http://dx.doi.org/10.1103/PhysRevD.90.105031}{\emph{Phys. Rev.} {\bf
  D90} (2014) 105031}, [\href{http://arxiv.org/abs/1210.3605}{{\tt
  1210.3605}}].

\bibitem{Kim:2012qf}
H.-C. Kim, J.~Kim and S.~Kim, \emph{{Instantons on the 5-sphere and
  M5-branes}},  \href{http://arxiv.org/abs/1211.0144}{{\tt 1211.0144}}.

\bibitem{Kim:2012ava}
H.-C. Kim and S.~Kim, \emph{{M5-branes from gauge theories on the 5-sphere}},
  \href{http://dx.doi.org/10.1007/JHEP05(2013)144}{\emph{JHEP} {\bf 05} (2013)
  144}, [\href{http://arxiv.org/abs/1206.6339}{{\tt 1206.6339}}].

\bibitem{Hollowood:2003cv}
T.~J. Hollowood, A.~Iqbal and C.~Vafa, \emph{{Matrix models, geometric
  engineering and elliptic genera}},
  \href{http://dx.doi.org/10.1088/1126-6708/2008/03/069}{\emph{JHEP} {\bf 03}
  (2008) 069}, [\href{http://arxiv.org/abs/hep-th/0310272}{{\tt
  hep-th/0310272}}].

\bibitem{Hohenegger:2013ala}
S.~Hohenegger and A.~Iqbal, \emph{{M-strings, elliptic genera and $\mathcal{N}
  = 4$ string amplitudes}},
  \href{http://dx.doi.org/10.1002/prop.201300035}{\emph{Fortsch. Phys.} {\bf
  62} (2014) 155--206}, [\href{http://arxiv.org/abs/1310.1325}{{\tt
  1310.1325}}].

\bibitem{Hohenegger:2015btj}
S.~Hohenegger, A.~Iqbal and S.-J. Rey, \emph{{Instanton-monopole correspondence
  from M-branes on $\mathbb S^1$ and little string theory}},
  \href{http://dx.doi.org/10.1103/PhysRevD.93.066016}{\emph{Phys. Rev.} {\bf
  D93} (2016) 066016}, [\href{http://arxiv.org/abs/1511.02787}{{\tt
  1511.02787}}].

\bibitem{Gadde:2009kb}
A.~Gadde, E.~Pomoni, L.~Rastelli and S.~S. Razamat, \emph{{S-duality and 2d
  Topological QFT}},
  \href{http://dx.doi.org/10.1007/JHEP03(2010)032}{\emph{JHEP} {\bf 03} (2010)
  032}, [\href{http://arxiv.org/abs/0910.2225}{{\tt 0910.2225}}].

\bibitem{2003math......6164N}
A.~{Narukawa}, \emph{The modular properties and the integral representations of
  the multiple elliptic gamma functions}, {\emph{ArXiv Mathematics e-prints}
  (June, 2003) }, [\href{http://arxiv.org/abs/math/0306164}{{\tt
  math/0306164}}].

\bibitem{Spiridonov:2013zma}
V.~P. Spiridonov, \emph{{Aspects of elliptic hypergeometric functions}},  in
  \emph{{RMS-Lecture Notes Series, no. 20 (2013), 347-361}}, 2013.
\newblock \href{http://arxiv.org/abs/1307.2876}{{\tt 1307.2876}}.

\bibitem{Felder200044}
G.~Felder and A.~Varchenko, \emph{The elliptic gamma function and
  sl(3, z)⋉z3},
  \href{http://dx.doi.org/http://dx.doi.org/10.1006/aima.2000.1951}{\emph{Advances
  in Mathematics} {\bf 156} (2000) 44 -- 76}.

\bibitem{DiPietro:2014bca}
L.~Di~Pietro and Z.~Komargodski, \emph{{Cardy formulae for SUSY theories in $d
  =$ 4 and $d =$ 6}},
  \href{http://dx.doi.org/10.1007/JHEP12(2014)031}{\emph{JHEP} {\bf 12} (2014)
  031}, [\href{http://arxiv.org/abs/1407.6061}{{\tt 1407.6061}}].

\bibitem{DiPietro:2016ond}
L.~Di~Pietro and M.~Honda, \emph{{Cardy Formula for 4d SUSY Theories and
  Localization}}, \href{http://dx.doi.org/10.1007/JHEP04(2017)055}{\emph{JHEP}
  {\bf 04} (2017) 055}, [\href{http://arxiv.org/abs/1611.00380}{{\tt
  1611.00380}}].

\bibitem{Ardehali:2015bla}
A.~Arabi~Ardehali, \emph{{High-temperature asymptotics of supersymmetric
  partition functions}},
  \href{http://dx.doi.org/10.1007/JHEP07(2016)025}{\emph{JHEP} {\bf 07} (2016)
  025}, [\href{http://arxiv.org/abs/1512.03376}{{\tt 1512.03376}}].

\bibitem{Ardehali:2015hya}
A.~Arabi~Ardehali, J.~T. Liu and P.~Szepietowski, \emph{{High-Temperature
  Expansion of Supersymmetric Partition Functions}},
  \href{http://dx.doi.org/10.1007/JHEP07(2015)113}{\emph{JHEP} {\bf 07} (2015)
  113}, [\href{http://arxiv.org/abs/1502.07737}{{\tt 1502.07737}}].

\bibitem{Clavelli:1973uk}
L.~Clavelli and J.~A. Shapiro, \emph{{Pomeron factorization in general dual
  models}}, \href{http://dx.doi.org/10.1016/0550-3213(73)90113-2}{\emph{Nucl.
  Phys.} {\bf B57} (1973) 490--535}.

\bibitem{Mironov:2016cyq}
A.~Mironov, A.~Morozov and Y.~Zenkevich, \emph{{Spectral duality in elliptic
  systems, six-dimensional gauge theories and topological strings}},
  \href{http://dx.doi.org/10.1007/JHEP05(2016)121}{\emph{JHEP} {\bf 05} (2016)
  121}, [\href{http://arxiv.org/abs/1603.00304}{{\tt 1603.00304}}].

\end{thebibliography}\endgroup

\end{document}